\documentclass[letter,prb,reprint]{revtex4-2}
\usepackage{graphicx}% Include figure files
\usepackage{bm}% bold math
\usepackage{amsmath,amsfonts,amssymb,braket,mathtools}
\usepackage[dvipdfmx]{hyperref}
\usepackage{color}
\hypersetup{
	colorlinks=true,
	linkcolor=blue,
	citecolor=green,
	bookmarks=true,
	bookmarksnumbered=true,
	pdfborder={0 0 0},
	bookmarkstype=toc
}
\usepackage{colortbl}
\begin{document}
\title{Finite-$q$ antiferrotoroidal and ferritoroidal order in a distorted kagome structure}
\author{Akimitsu Kirikoshi$^{1}$}
\author{Satoru Hayami$^{2}$}
\affiliation{
  $^{1}$Research Institute for Interdisciplinary Science, Okayama University, Okayama 700-8530, Japan
  \\
  $^{2}$Graduate School of Science, Hokkaido University, Hokkaido 060-0810, Japan
}

% \date{\today}

\begin{abstract}
  A highly geometrically frustrated lattice structure such as a distorted kagome (or quasikagome) structure enriches physical phenomena through coupling with the electronic structure, topology, and magnetism.
  Recently, it has been reported that an intermetallic HoAgGe exhibits two distinct magnetic structures with the finite magnetic vector $\bm{q}=(1/3,1/3,0)$:
  One is the partially ordered state in the intermediate-temperature region, and the other is the kagome spin ice state in the lowest-temperature region.
  We theoretically elucidate that the former is characterized by antiferrotoroidal ordering, while the latter is characterized by ferritoroidal ordering based on the multipole representation theory, which provides an opposite interpretation to previous studies.
  We also show how antiferrotoroidal and ferritoroidal orderings are microscopically formed by quantifying the magnetic toroidal moment activated in a multiorbital system.
  As a result, we find that the degree of distortion for the kagome structure plays a significant role in determining the nature of antiferrotoroidal and ferritoroidal orderings, which brings about the crossover between the antiferro-type and the ferri-type distributions of the magnetic toroidal dipole. We confirm such a tendency by evaluating the linear magnetoelectric effect.
  Our analysis can be applied irrespective of lattice structures and magnetic vectors without annoying the cluster origin.
\end{abstract}

\maketitle

A magnetic toroidal dipole (MTD), generated by a vortex structure of magnetic dipoles, is one of the fundamental quantities characterizing the antiferromagnetic structure. 
Since the MTD breaks both spatial inversion and time-reversal symmetries, the cross-correlated phenomenon known as the magnetoelectric (ME) effect has attracted considerable attention, especially for antiferromagnetic insulators~\cite{PhysRevB.76.214404,Spaldin_JPCM.20.434204,Kopaev_Physics_Uspekhi.52.1111}, e.g., $\mathrm{Cr}_{2}\mathrm{O}_{3}$~\cite{JETPL.69.330}, $\mathrm{Ga}_{2-x}\mathrm{Fe}_{x}\mathrm{O}_{3}$~\cite{JETP.87.146,JPSJ.74.1419}, $\mathrm{LiCoPO}_{4}$~\cite{Nature.449.702,ncomms5796}, and $\mathrm{Ba}_{2}\mathrm{CoGe}_{2}\mathrm{O}_{7}$~\cite{PhysRevB.84.094421}.
In recent years, the MTD order in metals has been studied intensively~\cite{JPSJ.83.014703, PhysRevB.90.024432, JApplPhys.127.213905}.
For example, the current-induced magnetization has been observed in $\mathrm{UNi}_{4}\mathrm{B}$~\cite{JPSJ.87.033702} under antiferromagnetic ordering with the MTD moment.
Furthermore, transport phenomena characteristic of the magnetic toroidal (MT) metals have been explored in both theory and experiment, such as nonreciprocal currents~\cite{PhysRevResearch.2.043081,PhysRevB.105.155157,Suzuki_PhysRevB.105.075201} and nonlinear Hall effects~\cite{PhysRevLett.127.277201,PhysRevLett.127.277202,PhysRevB.107.155109,ota2022zerofield}.
On the other hand, antiferroic-MTD order has attracted less attention since the cancellation of physical phenomena is anticipated.
Indeed, no prototype materials have been reported so far.

\begin{figure}[t]
  \centering
  \includegraphics[width=\linewidth]{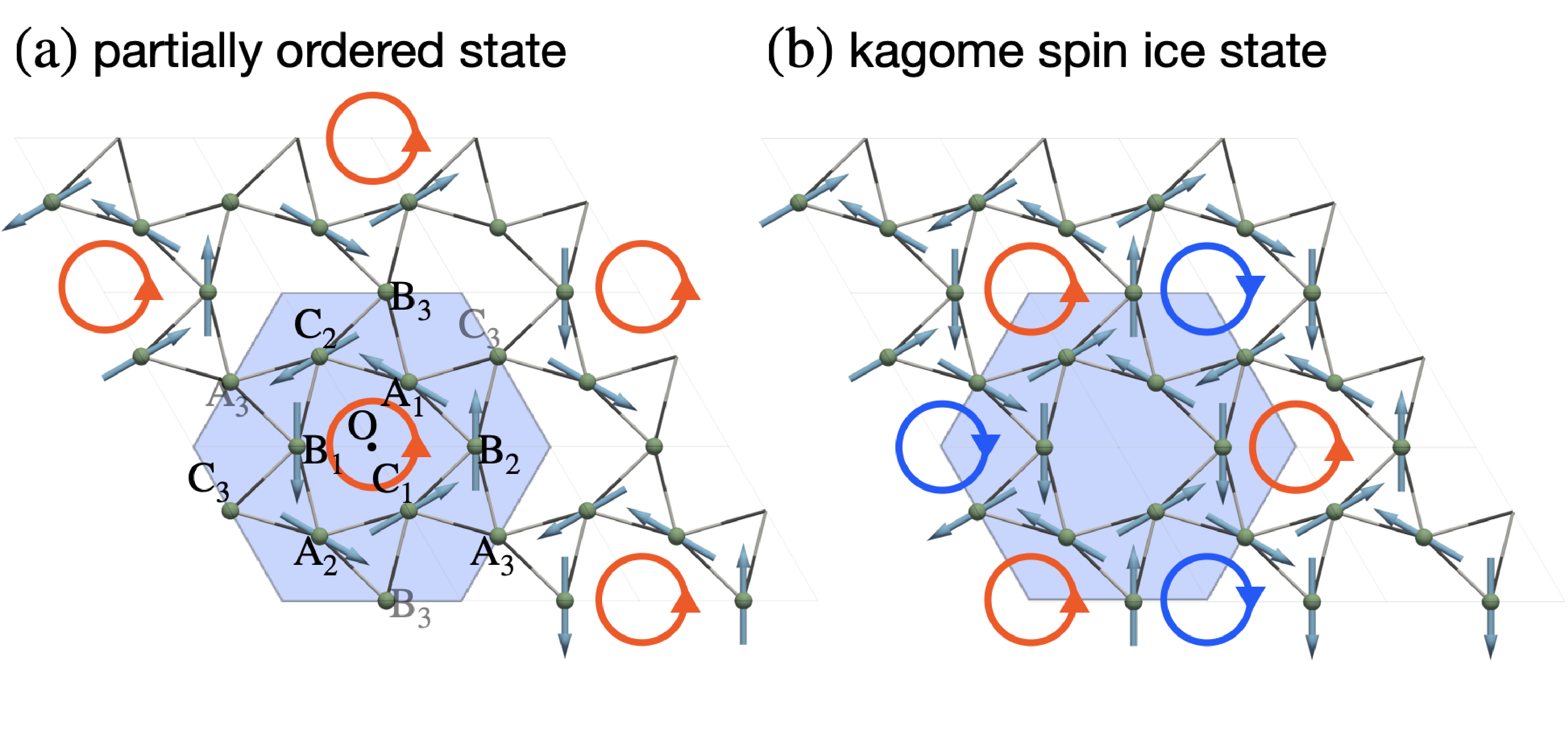}
  \caption{\label{fig:single_q_MT} Magnetic structures of (a) the partially ordered state and (b) the kagome spin ice state observed in $\mathrm{HoAgGe}$~\cite{science.aaw1666}.
  The blue hexagon represents a magnetic ($\sqrt{3}\times\sqrt{3}$) unit cell with the origin $\mathrm{O}$, and $\mathrm{P}_{1}-\mathrm{P}_{3}$ with $\mathrm{P}=\mathrm{A,B,C}$ are the labels of nine independent sites in the magnetic unit cell.
  The red counterclockwise (blue clockwise) arrows show a positive (negative) out-of-plane MTD moment defined by Eq.~\eqref{eq:site_cluster_MTD} at the center of each hexagon with six ordered magnetic moments.
  } 
\end{figure}

Recently, the rare-earth compound HoAgGe has been proposed as a candidate for a new antiferro MT metal, which exhibits a $\bm{q}=(1/3,1/3,0)$ magnetic structure in a $\sqrt{3}\times\sqrt{3}$ magnetic unit cell~\cite{science.aaw1666}.
By decreasing the temperature from the paramagnetic state under No. $189\ P\bar{6}2m$ ($D_{\mathrm{3h}}^{3}$) symmetry, this compound first shows the transition into a partially ordered state below $12$~K, where $2/3$ of $\mathrm{Ho}^{3+}$ ions exhibit a vortex spin structure, and the remaining $1/3$ of $\mathrm{Ho}^{3+}$ ions are disordered, as shown in Fig.~\ref{fig:single_q_MT}(a).
When the temperature is further decreased to $7
$~K, all the $\mathrm{Ho}^{3+}$ ions exhibit a complicated vortex spin structure, as shown in Fig.~\ref{fig:single_q_MT}(b).
Since this state consists of triangles with a two-in/one-out or one-in/two-out configuration, it is referred to as a kagome spin ice state.
Both magnetic phases are characterized by the same magnetic space group, $P\bar{6}^{\prime}m2^{\prime}$~\cite{science.aaw1666}.
Since the MTD belongs to the identity irreducible representation under $P\bar{6}^{\prime}m2^{\prime}$~\cite{PhysRevB.104.054412}, this compound can exhibit MTD-related physical phenomena even for antiferrotoroidal and ferritoroidal ordering, as detailed below.

The MTD moment has been defined in conventional form as~\cite{science.aaw1666}
\begin{equation}
  \bm{T}^{(\mathrm{c})}=\sum_{i}\bm{r}_{i}\times\bm{M}_{i},
  \label{eq:site_cluster_MTD}
\end{equation}
where $\bm{M}_{i}$ is the magnetic moment at the position $\bm{r}_{i}$ and the superscript $(\mathrm{c})$ stands for the cluster quantity.
When the cluster origin is set at the center of each hexagon, and the summation is taken for six sites on the hexagon as $\mathrm{A}_{1}$-$\mathrm{C}_{2}$-$\mathrm{B}_{1}$-$\mathrm{A}_{2}$-$\mathrm{C}_{1}$-$\mathrm{B}_{2}$ in Fig.~\ref{fig:single_q_MT}(a), one obtains the spatial distribution of the out-of-plane MTD moment as shown by the clockwise or counterclockwise arrows in Figs.~\ref{fig:single_q_MT}(a) and \ref{fig:single_q_MT}(b).
This intuitive illustration of the MTD distribution indicates that the partially ordered state corresponds to ferritoroidal ordering, whereas the kagome spin ice state corresponds to antiferrotoroidal ordering~\cite{science.aaw1666}.
In contrast to such a simple picture, the evaluation of the MTD moment using Eq.~\eqref{eq:site_cluster_MTD} is ill defined because the six sites consisting of the hexagon are not symmetry related to each other owing to the lack of sixfold rotational symmetry.
Moreover, the cluster origin is arbitrarily taken within Eq.~\eqref{eq:site_cluster_MTD}.

In this Letter, we investigate how to describe the MTD distribution under finite-$q$ magnetic structures without ambiguity based on the microscopic multipole representation theory~\cite{JPSJ.93.072001, PhysRevB.107.195118}.
We show that the partially ordered and kagome spin ice states are expressed as antiferrotoroidal and ferritoroidal orderings, respectively, which is opposed to the previous interpretation~\cite{science.aaw1666}. 
Then, we analyze a fundamental multiorbital model to examine the spatial distribution of the MTD. 
We find that the degree of distortion for the kagome structure affects the behavior of the MTD;
we show that the antiferro (ferri) MT state in the perfect kagome structure turns into the ferri (antiferro) MT state in the distorted kagome structure by changing the distortion parameter. 

\begin{figure}[t]
  \centering
  \includegraphics[width=\linewidth]{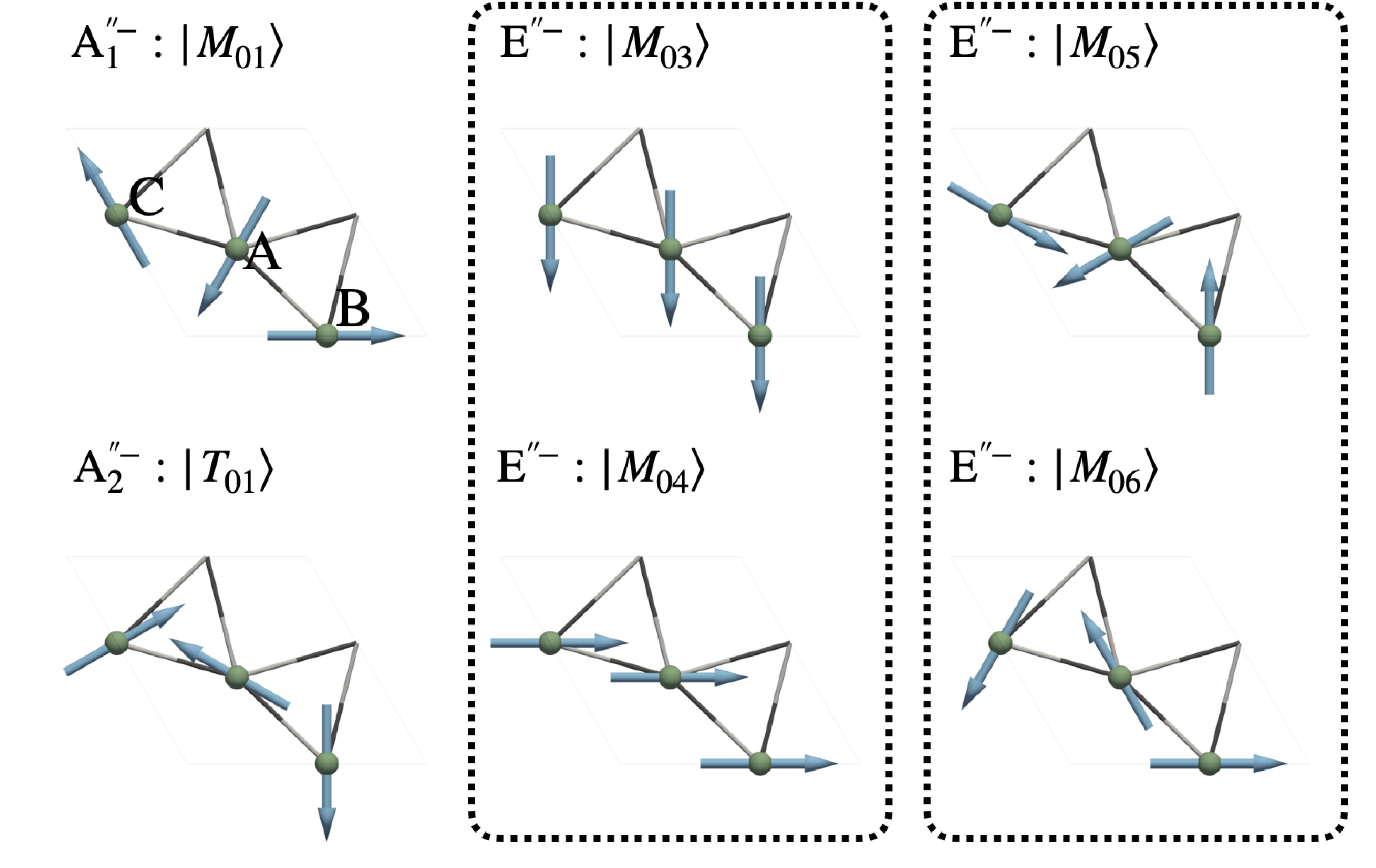}
  \caption{\label{fig:3_sub}
  SAMBs and their irreducible representations for the three-sublattice in-plane magnetic structure under the $D_{\mathrm{3h}}$ symmetry of the distorted kagome structure, where the superscript ``$-$'' represents odd parity for the time-reversal operation.
  The dashed lines enclose the bases for the two-dimensional irreducible representation $\mathrm{E}^{\prime\prime}$.
  }
\end{figure}

First, let us show how to describe antiferrotoroidal and ferritoroidal orderings under the finite-$q$ magnetic structure by adopting the symmetry-adapted multipole basis (SAMB)~\cite{PhysRevB.107.195118}, which gives the symmetry-related correspondence between the multipole and electronic degrees of freedom under the $\bm{q}=\bm{0}$ structure independent of the choice of the cluster origin.
Using the SAMB combined with the virtual cluster method, any three-sublattice magnetic structures can be expressed as a linear combination of six magnetic multipole bases $M_{i}$ and three MT multipole bases $T_{i}$.
Among them, the bases describing the in-plane spin modulations relevant to HoAgGe can be described by 
$\{\ket{M_{01}}, \ket{T_{01}}, \ket{M_{03}}, \ket{M_{04}}, \ket{M_{05}}, \ket{M_{06}}\}$, whereas the bases describing the out-of-plane spin modulations $\{\ket{M_{02}},\ket{T_{02}},\ket{T_{03}}\}$ are irrelevant in this study.
We show the spin configurations corresponding to six bases in Fig.~\ref{fig:3_sub}. 
The basis $\ket{T_{01}}$ corresponds to the MTD moment along the $z$ direction, i.e., $T_{z}$, from the symmetry.

The $\sqrt{3}\times \sqrt{3}$ magnetic structures in Fig.~\ref{fig:single_q_MT} can be constructed by taking the direct product of the spatial modulation with $\bm{q}=(1/3,1/3,0)$ and the three-sublattice SAMBs in Fig.~\ref{fig:3_sub}~\cite{PhysRevB.107.014407}.
\begin{subequations}
  \label{eq:single_q}
  The basis corresponding to the partially ordered state in Fig.~\ref{fig:single_q_MT}(a) is expressed by 
  \begin{equation}
    \ket{\mathrm{PO}}=\frac{2}{3}\left(\ket{T_{01}}\sin{\tilde{q}_{-}}+\ket{T_{01}^{\prime}}\sin{\tilde{q}_{+}}\right),
    \label{eq:partially_ordered}
  \end{equation}
  whereas that for the kagome spin ice state is expressed as 
  \begin{equation}
    \ket{\mathrm{SI}}=
    \frac{1}{3}\ket{T_{01}}-\frac{4}{3\sqrt{3}}
    (\ket{T_{01}}\cos{\tilde{q}_{-}}+\ket{T_{01}^{\prime}}\cos{\tilde{q}_{+}}),
    \label{eq:kagome_spin_ice}
  \end{equation}
  with $\ket{T_{01}^{\prime}}=\frac{1}{2}(\ket{M_{03}}-\ket{M_{05}})+\frac{\sqrt{3}}{2}(\ket{M_{04}}-\ket{M_{06}})$, and $\tilde{q}_{\pm}=\bm{q}\cdot\bm{r} \pm \pi/6$ for $\bm{q}=(1/3,1/3,0)$.
\end{subequations}
It is noted that this finite-$q$ expression is unique irrespective of the cluster origin.
The correspondence between the other $\sqrt{3}\times \sqrt{3}$ magnetic structures and multipoles is shown in Supplemental Material (SM)~\cite{Supplemental_Material}. 

The expression in Eq.~\eqref{eq:partially_ordered} indicates that the magnetic structure of the partially ordered state is characterized by the density wave of the MTD, meaning no $\bm{q}=\bm{0}$ component. 
In other words, the partially ordered state is regarded as the antiferrotoroidal ordering. 
On the other hand, the magnetic structure of the kagome spin ice state in Eq.~\eqref{eq:kagome_spin_ice} includes the $\bm{q}=\bm{0}$ component as well as the finite-$q$ one, which indicates the ferritoroidal ordering. 
Thus, the symmetry-adapted argument suggests an interpretation that contrasts with the previous intuitive interpretation~\cite{science.aaw1666}. 
Indeed, recent experiments have shown that the nonlinear conductivity originating from the MTD becomes large in the kagome spin ice state with the ferritoroidal structure, while that is negligibly small in the partially ordered state with the antiferrotoroidal structure~\cite{miyamoto}.

\begin{figure*}
  \centering
  \includegraphics[width=\linewidth]{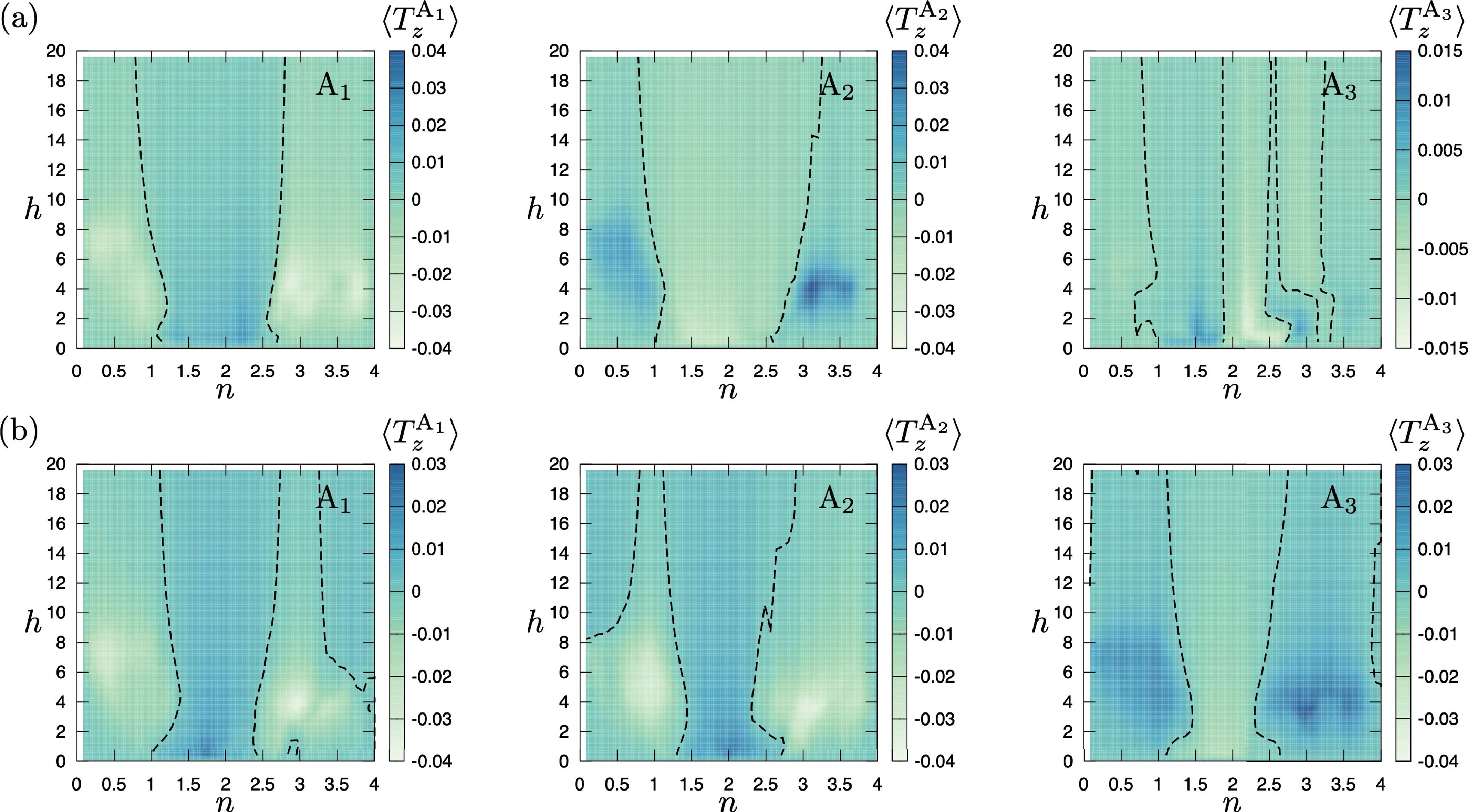}
  \caption{\label{fig:MT_distribution}The contour plot of the expectation values of the MTD $T_{z}^{\eta}$ at each site in the filling $n$ and molecular field $h$ plane under (a) the partially ordered state and (b) the kagome spin ice state with $\delta=0.16$.
  The dashed lines indicate that the expectation values of the MTD vanish.
  }
\end{figure*}
Next, we quantify the MTD under the magnetic structures in Fig.~\ref{fig:single_q_MT}.
We here consider the multiorbital model with different spatial parities to activate the MTD defined as the atomic-scale parity-mixing hybridization, e.g., $s$-$p$ and $p$-$d$ hybridizations.
Specifically, we adopt the periodic Anderson model under the two-dimensional distorted kagome structure, which is given by
\begin{equation}
  \label{eq:Ham}
  \begin{aligned}
    \mathcal{H}=&\, \sum_{\bm{k}\eta\eta^{\prime}\sigma}\varepsilon_{\bm{k}\eta\eta^{\prime}}c_{\bm{k}\eta\sigma}^{\dag}c_{\bm{k}\eta^{\prime}\sigma}
    +\sum_{\bm{k}\eta\eta^{\prime}\sigma\sigma^{\prime}}
    (V_{\bm{k}\eta\eta^{\prime}}^{\sigma\sigma^{\prime}}c_{\bm{k}\eta\sigma}^{\dag}p_{\bm{k}\eta^{\prime}\sigma^{\prime}}+\mathrm{H.c.})
    \\
    &\, +h\sum_{\bm{k}\eta\sigma\sigma^{\prime}}T_{\eta}^{\sigma\sigma^{\prime}}p_{\bm{k}\eta\sigma}^{\dag}p_{\bm{k}\eta\sigma^{\prime}},
  \end{aligned}
\end{equation}
where $c_{\bm{k}\eta\sigma}^{\dag}(c_{\bm{k}\eta\sigma})$ is the creation (annihilation) operator for the itinerant electron with momentum $\bm{k}$, sublattice $\eta
\in\{\mathrm{A}_{1},\mathrm{A}_{2},\mathrm{A}_{3},\mathrm{B}_{1},\mathrm{B}_{2},\mathrm{B}_{3},\mathrm{C}_{1},\mathrm{C}_{2},\mathrm{C}_{3}\}$, and spin $\sigma=\pm 1$; $p_{\bm{k}\eta\sigma}^{\dag}(p_{\bm{k}\eta\sigma})$ is that for the localized electron($\sigma$ represents the pseudospin).
The sublattices $(\mathrm{A}_{i},\mathrm{B}_{i},\mathrm{C}_{i})$ for $i=1,2,3$ are connected by the three-fold rotation to each other, whereas $(\mathrm{P}_{1},\mathrm{P}_{2},\mathrm{P}_{3})$ for $\mathrm{P}=\mathrm{A},\mathrm{B},\mathrm{C}$ are connected by the translation (see Fig.~\ref{fig:single_q_MT}).
The first term is the hopping term between the itinerant electrons
\begin{equation}
  \varepsilon_{\bm{k}\eta\eta^{\prime}}=t_{s}e^{i\bm{k}\cdot(\bm{r}_{\eta}-\bm{r}_{\eta^{\prime}})},
\end{equation}
where $t_{s}$ is the amplitude for the nearest-neighbor hopping and $\bm{r}_{\eta}$ is the position of the sublattice $\eta$ in Fig.~\ref{fig:single_q_MT}.
The second term is the hopping between the itinerant electron at $\eta$ with spin $\sigma$ and the localized one at $\eta^{\prime}$ with pseudospin $\sigma^{\prime}$.
The coefficients are expressed as 
\begin{equation}
  V_{\bm{k}\eta\eta^{\prime}}^{\sigma\sigma^{\prime}}=t_{sp}e^{i\bm{k}\cdot(\bm{r}_{\eta}-\bm{r}_{\eta^{\prime}})}w_{\eta\eta^{\prime}}^{\sigma\sigma^{\prime}},
\end{equation}
where $t_{sp}$ is the amplitude for the nearest-neighbor hopping and $w_{\eta\eta^{\prime}}^{\sigma\sigma^{\prime}}$ is the overlap integral~\cite{Supplemental_Material}.
We suppose the orbital degrees for the itinerant and localized electrons are $s$ and $p$ orbitals, respectively, for simplicity, where we set the local crystalline electric field for the $p$ (localized) orbital so that the magnetic easy axis is directed perpendicular to the mirror plane at each lattice site, as found in HoAgGe; see the detail in SM~\cite{Supplemental_Material}.
The third term represents the molecular field with the strength $h$ to induce the partially ordered state or kagome spin ice state, which originates from the mean-field decoupling of the Coulomb interaction.

It is noteworthy that the degree of distortion from the perfect kagome structure affects the hopping between itinerant and localized electrons, reflecting the anisotropy of the localized orbital.
To investigate the effect of the distortion on the MTD, we introduce the parameter $\delta$ as $\frac{1-\delta}{2}a(\frac{1}{2},\frac{\sqrt{3}}{2})$ ($a$ is the lattice constant of the unit cell in Fig.~\ref{fig:single_q_MT}) for the $\mathrm{A}_{1}$ site; the structure with $\delta=0$ corresponds to the perfect kagome structure, and that with $\delta=0.16$ corresponds to the distorted kagome structure in HoAgGe.
We also show the coordinates for other sublattices in SM~\cite{Supplemental_Material}.
In the following calculation, we fix $t_{s}=-1,t_{sp}=0.5$, $a=1$, and the temperature $T=10^{-3}$.

The atomic-scale operator of the $z$-component  MTD per sublattice $\eta$ is given by 
\begin{equation}
  T_{z}^{\eta}=\frac{1}{N_{\bm{k}}}\sum_{\bm{k}\sigma}(ic_{\bm{k}\eta\sigma}^{\dag}p_{\bm{k}\eta\sigma}+\mathrm{H.c.}),
  \label{eq:Tz_operator}
\end{equation}
where $N_{\bm{k}}$ is the number of $\bm{k}$-mesh in the the Brillouin zone; we set $N_{\bm{k}}=718^{2}$.
Owing to the three-fold rotation around the $z$-axis for both the partially ordered and kagome spin ice states, the expectation values of $T_{z}^{\eta}$, $\braket{T_{z}^{\eta}}$, at $\eta=\mathrm{A}_{i},\mathrm{B}_{i},\mathrm{C}_{i}$ are equivalent.

\begin{figure*}
  \centering
  \includegraphics[width=\linewidth]{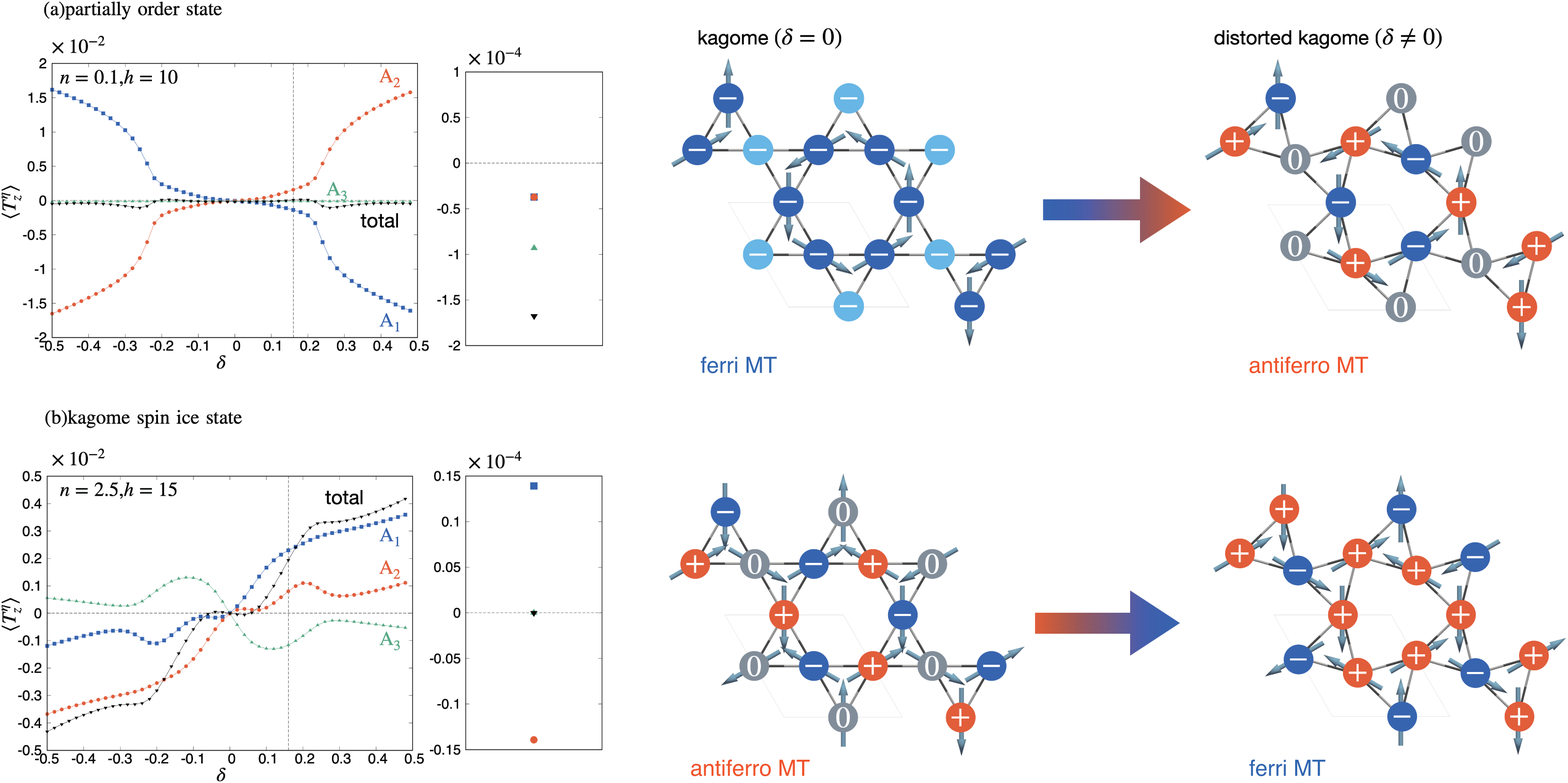}
  \caption{\label{fig:MT_distortion}
  Distortion parameter ($\delta$) dependence of the MTD in Eq.~\eqref{eq:Tz_operator} under (a) the partially ordered state and (b) the kagome spin ice state.
  The black points labeled as ``total'' are the rsults for $\braket{T_{1}^{\mathrm{A}_{1}}}+\braket{T_{1}^{\mathrm{A}_{2}}}+\braket{T_{1}^{\mathrm{A}_{3}}}$.
  The vertical dotted line at $\delta=0.16$ indicates the case of HoAgGe.
  The second panels show the results at $\delta=0$.
  The right panels show the schematic pictures of the change in the spatial distribution of the MTD, where the signs on each site represent the direction of the MTD.
  }
\end{figure*}
Figures~\ref{fig:MT_distribution}(a) and~\ref{fig:MT_distribution}(b) show $\braket{T_{z}^{\eta}}$ for the partially ordered and kagome spin ice states, respectively, against the electron filling $n$ and $h$.
In the partially ordered state, the MTDs at the ordered sites ($\mathrm{A}_{1}$ and $\mathrm{A}_{2}$) are antiferroic, whereas that at the disordered sites ($\mathrm{A}_3$) is relatively suppressed. 
This result means the antiferroic relation of the MTD satisfying $\braket{T_{z}^{\mathrm{A}_{1}}}=-\braket{T_{z}^{\mathrm{A}_{2}}}$ and $\braket{T_{z}^{\mathrm{A}_{3}}} \simeq 0$, as shown in Fig.~\ref{fig:MT_distribution}(a).
On the other hand, in the kagome spin ice state, the MTD at the $\mathrm{A}_{1}$ and $\mathrm{A}_{2}$ sites with the same magnetic moment shows a ferroic behavior over a wide parameter range, while the MTD at the $\mathrm{A}_{3}$ site with the opposite magnetic moment tends to be developed in the opposite direction.
This tendency of $\braket{T_{z}^{\mathrm{A}_{1}}}\simeq \braket{T_{z}^{\mathrm{A}_{2}}}\simeq -\braket{T_{z}^{\mathrm{A}_{3}}}$ holds for a wide parameter range, as shown in Fig.~\ref{fig:MT_distribution}(b).
These results are consistent with our symmetry-based interpretation of the antiferrotoroidal and ferritoroidal orderings.

Meanwhile, the relations of the $\braket{T_{z}^{\mathrm{A}_{1}}}=-\braket{T_{z}^{\mathrm{A}_{2}}}$ and $\braket{T_{z}^{\mathrm{A}_{3}}}=0$ in the partially ordered state and $\braket{T_{z}^{\mathrm{A}_{1}}}=\braket{T_{z}^{\mathrm{A}_{2}}}=-\braket{T_{z}^{\mathrm{A}_{3}}}$ in the kagome spin ice state expected from the cluster multipole theory are violated in some parameter range.
In order to understand the origin of such a deviation, we investigate the $\delta$ dependence of $\braket{T_{z}^{\eta}}$ in the left panel of Fig.~\ref{fig:MT_distortion}.
We find that the antiferro MT and ferri MT natures are transformed to each other by changing $\delta$. 
For the perfect kagome structure with $\delta=0$, we obtain the ferri(antiferro)-type distribution of the MTD, i.e., $\braket{T_{z}^{\mathrm{A}_{1}}}=\braket{T_{z}^{\mathrm{A}_{2}}}>0$ and $\braket{T_{z}^{\mathrm{A}_{3}}}\neq0$ ($\braket{T_{z}^{\mathrm{A}_{1}}}=-\braket{T_{z}^{\mathrm{A}_{2}}}$ and $\braket{T_{z}^{\mathrm{A}_{3}}}=0$), rather than the antiferro(ferri)-type one in the partially ordered (kagome spin ice) state. 
This behavior for $\delta=0$ is also understood from the SAMB under the perfect kagome lattice structure, as shown in SM~\cite{Supplemental_Material}.

When $\delta$ is nonzero, the MTD at each site changes so that its distribution in the partially ordered (kagome spin ice) state changes from the ferri (antiferro) MT distribution to the antiferro (ferri) MT  one in a crossover way.
By analyzing the power dependence of $\delta$ in terms of $\braket{T_{z}^{\eta}}$~\cite{JPSJ.91.014701}, we find that the even-order $O(\delta^{2n})$ contributes as $\braket{T_{z}^{\mathrm{A}_{1}}}=\braket{T_{z}^{\mathrm{A}_{2}}}\neq \braket{T_{z}^{\mathrm{A}_{3}}}$, whereas the odd-order $O(\delta^{2n+1})$ contributes as $\braket{T_{z}^{\mathrm{A}_{1}}}=-\braket{T_{z}^{\mathrm{A}_{2}}}$ and $\braket{T_{z}^{\mathrm{A}_{3}}}=0$ in the partially ordered state.
Similarly, we find that the even-order $O(\delta^{2n})$ contributes as $\braket{T_{z}^{\mathrm{A}_{1}}}=-\braket{T_{z}^{\mathrm{A}_{2}}}\neq 0$ and $\braket{T_{z}^{\mathrm{A}_{3}}}=0$, whereas the odd-order $O(\delta^{2n+1})$ contributes as $\braket{T_{z}^{\mathrm{A}_{1}}}=\braket{T_{z}^{\mathrm{A}_{2}}}\neq 0$ and $\braket{T_{z}^{\mathrm{A}_{3}}}\neq 0$ in the kagome spin ice state.
We show the schematic picture of the crossover between the ferri MT and antiferro MT orders in the right panel of Fig.~\ref{fig:MT_distortion}.

Finally, we relate the MTD distribution with the linear ME response.
We evaluate the ME tensor defined by $M_{\alpha}=\chi_{\alpha\beta}E_{\beta}\ (\alpha,\beta=x,y,z)$ from the Kubo formula.
Since the magnetization can be decomposed into the sum of the contributions from each sublattice $M_{\alpha}=\sum_{\eta}M_{\alpha}^{\eta}$, the sublattice-dependent form is defined by 
\begin{equation}
  \chi_{\alpha\beta}^{\eta}=\frac{e\mu_{\mathrm{B}}\hbar}{iN_{\bm{k}}}\sum_{\bm{k}}\sum_{n\neq m}\frac{f_{n\bm{k}}-f_{m\bm{k}}}{(\varepsilon_{n\bm{k}}-\varepsilon_{m\bm{k}})^{2}}\mu_{\alpha\bm{k}}^{\eta,nm}v_{\beta\bm{k}}^{mn},
\end{equation}
where $e>0$ is the elementary charge, $\mu_{\mathrm{B}}$ is the Bohr magneton, $\hbar$ is the Dirac constant, $\varepsilon_{n\bm{k}}$ is the band energy with the index $n$, and $f_{n\bm{k}}$ is the Fermi--Dirac distribution function. 
$\mu_{\alpha\bm{k}}^{\eta,nm}=\braket{n\bm{k}|\mu_{\alpha}^{\eta}|m\bm{k}}$ with $\mu_{\alpha}^{\eta}=l_{\alpha}^{\eta}+2s_{\alpha}^{\eta}$ is the band representation of the magnetic moment operator at $\eta$ and $v_{\beta\bm{k}}^{mn}=\hbar^{-1}\braket{m\bm{k}|\partial_{\beta}h(\bm{k})|n\bm{k}}$ is that of the velocity operator.
We also define the total ME tensor as $\chi_{\alpha\beta}^{\mathrm{tot}}=\chi_{\alpha\beta}^{\mathrm{A}_{1}}+\chi_{\alpha\beta}^{\mathrm{A}_{2}}+\chi_{\alpha\beta}^{\mathrm{A}_{3}}$.
Since nonzero $T_{z}$ contributes to $\chi_{xy}=-\chi_{yx}$~\cite{PhysRevB.98.165110,PhysRevB.98.245129}, we evaluate the antisymmetric component of $\chi_{xy}$, that is, $\chi^{(\mathrm{A})}_{xy}=(\chi_{xy}-\chi_{yx})/2$ at each site.
We set $e=\hbar=\mu_{\mathrm{B}}=1$.

\begin{figure}
  \centering
  \includegraphics[width=\linewidth]{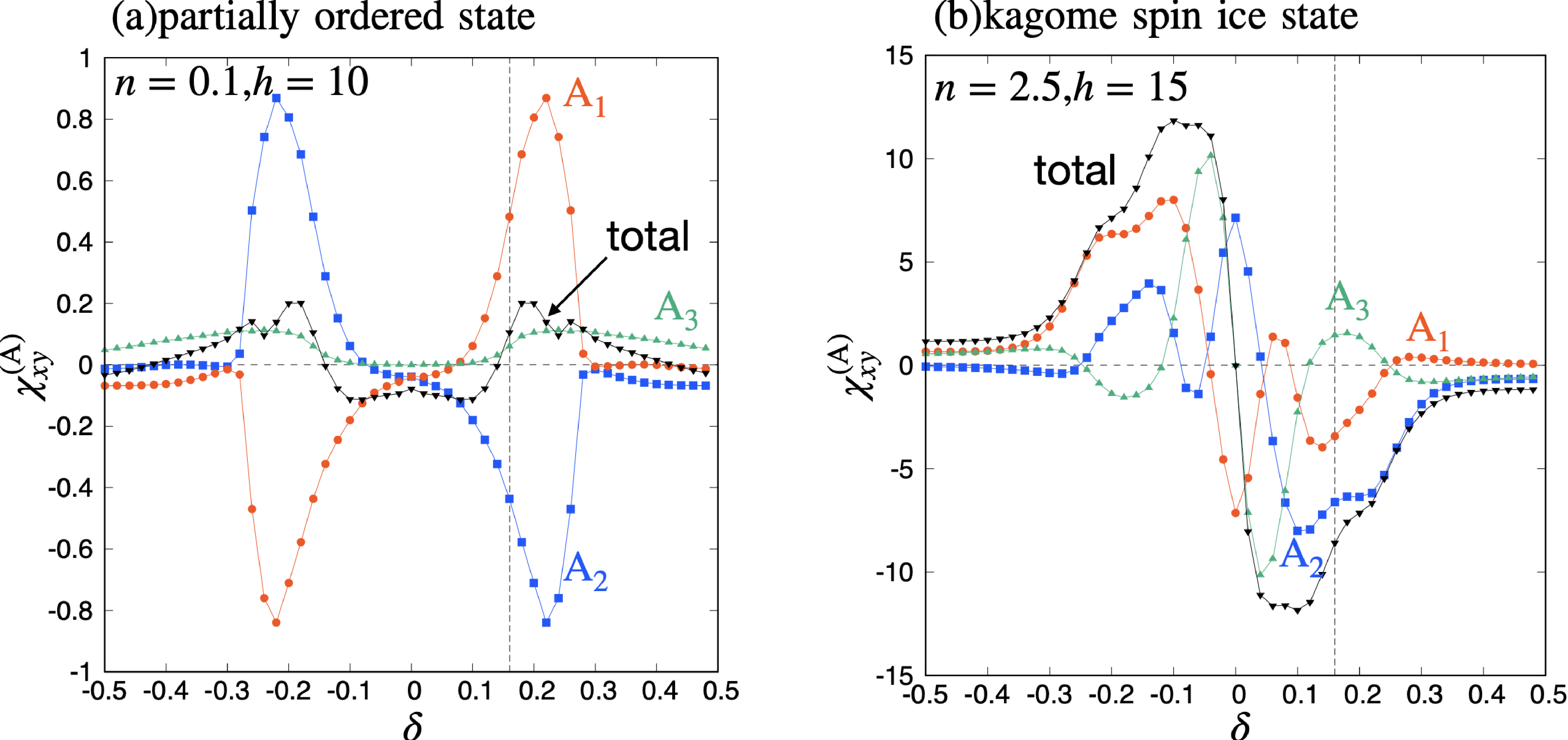}
  \caption{\label{fig:ME_distortion}
  Distortion parameter $(\delta)$ dependence of the ME tensor $\chi^{\mathrm(A)}_{xy}$ per site and the sum of them, under (a) the partially ordered state and (b) the kagome spin ice state.
  The vertical dotted line at $\delta=0.16$ indicates the case of HoAgGe.
  The parameters are the same as Fig.~\ref{fig:MT_distortion}.
  }
\end{figure}

Figure~\ref{fig:ME_distortion} shows the $\delta$ dependence of $\chi^{(\mathrm{A})}_{xy}$.
For the kagome structure with $\delta=0$, the total ME tensor is nonzero under the partially ordered state owing to the ferritoroidal MTD distribution.
When the distorted kagome structure is considered, the site-resolved ME effect becomes large for the  $\mathrm{A}_{1}$ and $\mathrm{A}_{2}$ sites, although they are canceled out with each other owing to the antiferrotoroidal MTD distribution, as shown in Fig.~\ref{fig:MT_distortion}(a). 
As a result, the total ME tensor is suppressed for all $\delta$ in the partially ordered state.
On the other hand, the ME tensor is enhanced in the kagome spin ice state under the lattice distortion, as shown in Fig.~\ref{fig:ME_distortion}(b), because the ferritoroidal MTD distribution with a net component is realized for $\delta \neq 0$.
We also confirmed that the power of $\delta$ dependence of $\chi_{xy}^{(\mathrm{A})}$ is the same as that of the MTD moment for both magnetic orderings.
In SM~\cite{Supplemental_Material}, we show the overall behavior of $\chi_{xy}^{(\mathrm{A})}$ for the filling and molecular field.

In summary, we theoretically investigated the foundations of the antiferrotoroidal and ferritoroidal orderings under the finite-$q$ magnetic structures.
In a previous study~\cite{science.aaw1666}, the spatial distribution of the local toroidal moment has been discussed based on Eq.~\eqref{eq:site_cluster_MTD}, and they conclude that the partially ordered (kagome spin ice) state has a net (no) MT moment.
However, applying the SAMBs to the two magnetic structures in $\mathrm{HoAgGe}$, we found that the partially ordered state is characterized by the antiferrotoroidal ordering.
In contrast, the kagome spin ice state is characterized by the ferritoroidal ordering with the uniform component.
By evaluating the spatial distribution of the atomic-scale MTD and linear ME tensor, we have shown that the lattice distortion parameter $\delta$ drives a crossover from the kagome structure-origin (even-power contribution of $\delta$) to the distorted kagome structure-origin (odd-power contribution of $\delta$).
When the latter contribution is primary, the partially ordered state and the kagome spin ice state realized in HoAgGe are well characterized by the antiferro and ferri MT states, respectively.
Our analysis will provide insight into the physical properties of the finite-$q$ MT ordering in $\mathrm{HoAgGe}$.

\begin{acknowledgments}
  The authors acknowledge T. Miyamoto, M. Shimozawa, S. Hosoi, and K. Izawa for fruitful discussions.
  This research was supported by JSPS KAKENHI Grants No. JP21H01037, No. JP22H00101, No. JP22H01183, No. JP23K03288, No. JP23H04869, and No. JP23K20827, by JST CREST (JPMJCR23O4), and by JST FOREST (JPMJFR2366).
  Parts of the numerical calculations were performed in the supercomputing systems in ISSP, the University of Tokyo.
\end{acknowledgments}

\bibliography{manuscript.bbl}

\begin{thebibliography}{30}%
\makeatletter
\providecommand \@ifxundefined [1]{%
 \@ifx{#1\undefined}
}%
\providecommand \@ifnum [1]{%
 \ifnum #1\expandafter \@firstoftwo
 \else \expandafter \@secondoftwo
 \fi
}%
\providecommand \@ifx [1]{%
 \ifx #1\expandafter \@firstoftwo
 \else \expandafter \@secondoftwo
 \fi
}%
\providecommand \natexlab [1]{#1}%
\providecommand \enquote  [1]{``#1''}%
\providecommand \bibnamefont  [1]{#1}%
\providecommand \bibfnamefont [1]{#1}%
\providecommand \citenamefont [1]{#1}%
\providecommand \href@noop [0]{\@secondoftwo}%
\providecommand \href [0]{\begingroup \@sanitize@url \@href}%
\providecommand \@href[1]{\@@startlink{#1}\@@href}%
\providecommand \@@href[1]{\endgroup#1\@@endlink}%
\providecommand \@sanitize@url [0]{\catcode `\\12\catcode `\$12\catcode `\&12\catcode `\#12\catcode `\^12\catcode `\_12\catcode `\%12\relax}%
\providecommand \@@startlink[1]{}%
\providecommand \@@endlink[0]{}%
\providecommand \url  [0]{\begingroup\@sanitize@url \@url }%
\providecommand \@url [1]{\endgroup\@href {#1}{\urlprefix }}%
\providecommand \urlprefix  [0]{URL }%
\providecommand \Eprint [0]{\href }%
\providecommand \doibase [0]{https://doi.org/}%
\providecommand \selectlanguage [0]{\@gobble}%
\providecommand \bibinfo  [0]{\@secondoftwo}%
\providecommand \bibfield  [0]{\@secondoftwo}%
\providecommand \translation [1]{[#1]}%
\providecommand \BibitemOpen [0]{}%
\providecommand \bibitemStop [0]{}%
\providecommand \bibitemNoStop [0]{.\EOS\space}%
\providecommand \EOS [0]{\spacefactor3000\relax}%
\providecommand \BibitemShut  [1]{\csname bibitem#1\endcsname}%
\let\auto@bib@innerbib\@empty
%</preamble>
\bibitem [{\citenamefont {Ederer}\ and\ \citenamefont {Spaldin}(2007)}]{PhysRevB.76.214404}%
  \BibitemOpen
  \bibfield  {author} {\bibinfo {author} {\bibfnamefont {C.}~\bibnamefont {Ederer}}\ and\ \bibinfo {author} {\bibfnamefont {N.~A.}\ \bibnamefont {Spaldin}},\ }\href@noop {} {\bibfield  {journal} {\bibinfo  {journal} {Phys. Rev. B}\ }\textbf {\bibinfo {volume} {76}},\ \bibinfo {pages} {214404} (\bibinfo {year} {2007})}\BibitemShut {NoStop}%
\bibitem [{\citenamefont {Spaldin}\ \emph {et~al.}(2008)\citenamefont {Spaldin}, \citenamefont {Fiebig},\ and\ \citenamefont {Mostovoy}}]{Spaldin_JPCM.20.434204}%
  \BibitemOpen
  \bibfield  {author} {\bibinfo {author} {\bibfnamefont {N.~A.}\ \bibnamefont {Spaldin}}, \bibinfo {author} {\bibfnamefont {M.}~\bibnamefont {Fiebig}},\ and\ \bibinfo {author} {\bibfnamefont {M.}~\bibnamefont {Mostovoy}},\ }\href@noop {} {\bibfield  {journal} {\bibinfo  {journal} {J. Phys. Condens. Matter}\ }\textbf {\bibinfo {volume} {20}},\ \bibinfo {pages} {434203} (\bibinfo {year} {2008})}\BibitemShut {NoStop}%
\bibitem [{\citenamefont {Kopaev}(2009)}]{Kopaev_Physics_Uspekhi.52.1111}%
  \BibitemOpen
  \bibfield  {author} {\bibinfo {author} {\bibfnamefont {Y.~V.}\ \bibnamefont {Kopaev}},\ }\href@noop {} {\bibfield  {journal} {\bibinfo  {journal} {Phys.-Uspekhi}\ }\textbf {\bibinfo {volume} {52}},\ \bibinfo {pages} {1111} (\bibinfo {year} {2009})}\BibitemShut {NoStop}%
\bibitem [{\citenamefont {Popov}\ \emph {et~al.}(1999)\citenamefont {Popov}, \citenamefont {Kadomtseva}, \citenamefont {Belov}, \citenamefont {Vorob'ev},\ and\ \citenamefont {Zvezdin}}]{JETPL.69.330}%
  \BibitemOpen
  \bibfield  {author} {\bibinfo {author} {\bibfnamefont {Y.~F.}\ \bibnamefont {Popov}}, \bibinfo {author} {\bibfnamefont {A.~M.}\ \bibnamefont {Kadomtseva}}, \bibinfo {author} {\bibfnamefont {D.~V.}\ \bibnamefont {Belov}}, \bibinfo {author} {\bibfnamefont {G.~P.}\ \bibnamefont {Vorob'ev}},\ and\ \bibinfo {author} {\bibfnamefont {A.~K.}\ \bibnamefont {Zvezdin}},\ }\href@noop {} {\bibfield  {journal} {\bibinfo  {journal} {J. Exp. Theor. Phys. Lett.}\ }\textbf {\bibinfo {volume} {69}},\ \bibinfo {pages} {330} (\bibinfo {year} {1999})}\BibitemShut {NoStop}%
\bibitem [{\citenamefont {Popov}\ \emph {et~al.}(1998)\citenamefont {Popov}, \citenamefont {Kadomtseva}, \citenamefont {Vorob'ev}, \citenamefont {Timofeeva}, \citenamefont {Ustinin}, \citenamefont {Zvezdin},\ and\ \citenamefont {Tegeranchi}}]{JETP.87.146}%
  \BibitemOpen
  \bibfield  {author} {\bibinfo {author} {\bibfnamefont {Y.~F.}\ \bibnamefont {Popov}}, \bibinfo {author} {\bibfnamefont {A.~M.}\ \bibnamefont {Kadomtseva}}, \bibinfo {author} {\bibfnamefont {G.~P.}\ \bibnamefont {Vorob'ev}}, \bibinfo {author} {\bibfnamefont {V.~A.}\ \bibnamefont {Timofeeva}}, \bibinfo {author} {\bibfnamefont {D.~M.}\ \bibnamefont {Ustinin}}, \bibinfo {author} {\bibfnamefont {A.~K.}\ \bibnamefont {Zvezdin}},\ and\ \bibinfo {author} {\bibfnamefont {M.~M.}\ \bibnamefont {Tegeranchi}},\ }\href@noop {} {\bibfield  {journal} {\bibinfo  {journal} {J. Exp. Theor. Phys.}\ }\textbf {\bibinfo {volume} {87}},\ \bibinfo {pages} {146} (\bibinfo {year} {1998})}\BibitemShut {NoStop}%
\bibitem [{\citenamefont {Arima}\ \emph {et~al.}(2005)\citenamefont {Arima}, \citenamefont {Jung}, \citenamefont {Matsubara}, \citenamefont {Kubota}, \citenamefont {He}, \citenamefont {Kaneko},\ and\ \citenamefont {Tokura}}]{JPSJ.74.1419}%
  \BibitemOpen
  \bibfield  {author} {\bibinfo {author} {\bibfnamefont {T.-h.}\ \bibnamefont {Arima}}, \bibinfo {author} {\bibfnamefont {J.-H.}\ \bibnamefont {Jung}}, \bibinfo {author} {\bibfnamefont {M.}~\bibnamefont {Matsubara}}, \bibinfo {author} {\bibfnamefont {M.}~\bibnamefont {Kubota}}, \bibinfo {author} {\bibfnamefont {J.-P.}\ \bibnamefont {He}}, \bibinfo {author} {\bibfnamefont {Y.}~\bibnamefont {Kaneko}},\ and\ \bibinfo {author} {\bibfnamefont {Y.}~\bibnamefont {Tokura}},\ }\href@noop {} {\bibfield  {journal} {\bibinfo  {journal} {J. Phys. Soc. Jpn.}\ }\textbf {\bibinfo {volume} {74}},\ \bibinfo {pages} {1419} (\bibinfo {year} {2005})}\BibitemShut {NoStop}%
\bibitem [{\citenamefont {Van~Aken}\ \emph {et~al.}(2007)\citenamefont {Van~Aken}, \citenamefont {Rivera}, \citenamefont {Schmid},\ and\ \citenamefont {Fiebig}}]{Nature.449.702}%
  \BibitemOpen
  \bibfield  {author} {\bibinfo {author} {\bibfnamefont {B.~B.}\ \bibnamefont {Van~Aken}}, \bibinfo {author} {\bibfnamefont {J.-P.}\ \bibnamefont {Rivera}}, \bibinfo {author} {\bibfnamefont {H.}~\bibnamefont {Schmid}},\ and\ \bibinfo {author} {\bibfnamefont {M.}~\bibnamefont {Fiebig}},\ }\href@noop {} {\bibfield  {journal} {\bibinfo  {journal} {Nature}\ }\textbf {\bibinfo {volume} {449}},\ \bibinfo {pages} {702} (\bibinfo {year} {2007})}\BibitemShut {NoStop}%
\bibitem [{\citenamefont {Zimmermann}\ \emph {et~al.}(2014)\citenamefont {Zimmermann}, \citenamefont {Meier},\ and\ \citenamefont {Fiebig}}]{ncomms5796}%
  \BibitemOpen
  \bibfield  {author} {\bibinfo {author} {\bibfnamefont {A.~S.}\ \bibnamefont {Zimmermann}}, \bibinfo {author} {\bibfnamefont {D.}~\bibnamefont {Meier}},\ and\ \bibinfo {author} {\bibfnamefont {M.}~\bibnamefont {Fiebig}},\ }\href@noop {} {\bibfield  {journal} {\bibinfo  {journal} {Nat. Commun.}\ }\textbf {\bibinfo {volume} {5}},\ \bibinfo {pages} {4796} (\bibinfo {year} {2014})}\BibitemShut {NoStop}%
\bibitem [{\citenamefont {Toledano}\ \emph {et~al.}(2011)\citenamefont {Toledano}, \citenamefont {Khalyavin},\ and\ \citenamefont {Chapon}}]{PhysRevB.84.094421}%
  \BibitemOpen
  \bibfield  {author} {\bibinfo {author} {\bibfnamefont {P.}~\bibnamefont {Toledano}}, \bibinfo {author} {\bibfnamefont {D.~D.}\ \bibnamefont {Khalyavin}},\ and\ \bibinfo {author} {\bibfnamefont {L.~C.}\ \bibnamefont {Chapon}},\ }\href@noop {} {\bibfield  {journal} {\bibinfo  {journal} {Phys. Rev. B}\ }\textbf {\bibinfo {volume} {84}},\ \bibinfo {pages} {094421} (\bibinfo {year} {2011})}\BibitemShut {NoStop}%
\bibitem [{\citenamefont {Yanase}(2014)}]{JPSJ.83.014703}%
  \BibitemOpen
  \bibfield  {author} {\bibinfo {author} {\bibfnamefont {Y.}~\bibnamefont {Yanase}},\ }\href@noop {} {\bibfield  {journal} {\bibinfo  {journal} {J. Phys. Soc. Jpn.}\ }\textbf {\bibinfo {volume} {83}},\ \bibinfo {pages} {014703} (\bibinfo {year} {2014})}\BibitemShut {NoStop}%
\bibitem [{\citenamefont {Hayami}\ \emph {et~al.}(2014)\citenamefont {Hayami}, \citenamefont {Kusunose},\ and\ \citenamefont {Motome}}]{PhysRevB.90.024432}%
  \BibitemOpen
  \bibfield  {author} {\bibinfo {author} {\bibfnamefont {S.}~\bibnamefont {Hayami}}, \bibinfo {author} {\bibfnamefont {H.}~\bibnamefont {Kusunose}},\ and\ \bibinfo {author} {\bibfnamefont {Y.}~\bibnamefont {Motome}},\ }\href@noop {} {\bibfield  {journal} {\bibinfo  {journal} {Phys. Rev. B}\ }\textbf {\bibinfo {volume} {90}},\ \bibinfo {pages} {024432} (\bibinfo {year} {2014})}\BibitemShut {NoStop}%
\bibitem [{\citenamefont {Th{\"o}le}\ \emph {et~al.}(2020)\citenamefont {Th{\"o}le}, \citenamefont {Keliri},\ and\ \citenamefont {Spaldin}}]{JApplPhys.127.213905}%
  \BibitemOpen
  \bibfield  {author} {\bibinfo {author} {\bibfnamefont {F.}~\bibnamefont {Th{\"o}le}}, \bibinfo {author} {\bibfnamefont {A.}~\bibnamefont {Keliri}},\ and\ \bibinfo {author} {\bibfnamefont {N.~A.}\ \bibnamefont {Spaldin}},\ }\href@noop {} {\bibfield  {journal} {\bibinfo  {journal} {J. Appl. Phys.}\ }\textbf {\bibinfo {volume} {127}},\ \bibinfo {pages} {213905} (\bibinfo {year} {2020})}\BibitemShut {NoStop}%
\bibitem [{\citenamefont {Saito}\ \emph {et~al.}(2018)\citenamefont {Saito}, \citenamefont {Uenishi}, \citenamefont {Miura}, \citenamefont {Tabata}, \citenamefont {Hidaka}, \citenamefont {Yanagisawa},\ and\ \citenamefont {Amitsuka}}]{JPSJ.87.033702}%
  \BibitemOpen
  \bibfield  {author} {\bibinfo {author} {\bibfnamefont {H.}~\bibnamefont {Saito}}, \bibinfo {author} {\bibfnamefont {K.}~\bibnamefont {Uenishi}}, \bibinfo {author} {\bibfnamefont {N.}~\bibnamefont {Miura}}, \bibinfo {author} {\bibfnamefont {C.}~\bibnamefont {Tabata}}, \bibinfo {author} {\bibfnamefont {H.}~\bibnamefont {Hidaka}}, \bibinfo {author} {\bibfnamefont {T.}~\bibnamefont {Yanagisawa}},\ and\ \bibinfo {author} {\bibfnamefont {H.}~\bibnamefont {Amitsuka}},\ }\href@noop {} {\bibfield  {journal} {\bibinfo  {journal} {J. Phys. Soc. Jpn.}\ }\textbf {\bibinfo {volume} {87}},\ \bibinfo {pages} {033702} (\bibinfo {year} {2018})}\BibitemShut {NoStop}%
\bibitem [{\citenamefont {Watanabe}\ and\ \citenamefont {Yanase}(2020)}]{PhysRevResearch.2.043081}%
  \BibitemOpen
  \bibfield  {author} {\bibinfo {author} {\bibfnamefont {H.}~\bibnamefont {Watanabe}}\ and\ \bibinfo {author} {\bibfnamefont {Y.}~\bibnamefont {Yanase}},\ }\href@noop {} {\bibfield  {journal} {\bibinfo  {journal} {Phys. Rev. Res.}\ }\textbf {\bibinfo {volume} {2}},\ \bibinfo {pages} {043081} (\bibinfo {year} {2020})}\BibitemShut {NoStop}%
\bibitem [{\citenamefont {Yatsushiro}\ \emph {et~al.}(2022)\citenamefont {Yatsushiro}, \citenamefont {Oiwa}, \citenamefont {Kusunose},\ and\ \citenamefont {Hayami}}]{PhysRevB.105.155157}%
  \BibitemOpen
  \bibfield  {author} {\bibinfo {author} {\bibfnamefont {M.}~\bibnamefont {Yatsushiro}}, \bibinfo {author} {\bibfnamefont {R.}~\bibnamefont {Oiwa}}, \bibinfo {author} {\bibfnamefont {H.}~\bibnamefont {Kusunose}},\ and\ \bibinfo {author} {\bibfnamefont {S.}~\bibnamefont {Hayami}},\ }\href@noop {} {\bibfield  {journal} {\bibinfo  {journal} {Phys. Rev. B}\ }\textbf {\bibinfo {volume} {105}},\ \bibinfo {pages} {155157} (\bibinfo {year} {2022})}\BibitemShut {NoStop}%
\bibitem [{\citenamefont {Suzuki}(2022)}]{Suzuki_PhysRevB.105.075201}%
  \BibitemOpen
  \bibfield  {author} {\bibinfo {author} {\bibfnamefont {Y.}~\bibnamefont {Suzuki}},\ }\href@noop {} {\bibfield  {journal} {\bibinfo  {journal} {Phys. Rev. B}\ }\textbf {\bibinfo {volume} {105}},\ \bibinfo {pages} {075201} (\bibinfo {year} {2022})}\BibitemShut {NoStop}%
\bibitem [{\citenamefont {Wang}\ \emph {et~al.}(2021)\citenamefont {Wang}, \citenamefont {Gao},\ and\ \citenamefont {Xiao}}]{PhysRevLett.127.277201}%
  \BibitemOpen
  \bibfield  {author} {\bibinfo {author} {\bibfnamefont {C.}~\bibnamefont {Wang}}, \bibinfo {author} {\bibfnamefont {Y.}~\bibnamefont {Gao}},\ and\ \bibinfo {author} {\bibfnamefont {D.}~\bibnamefont {Xiao}},\ }\href@noop {} {\bibfield  {journal} {\bibinfo  {journal} {Phys. Rev. Lett.}\ }\textbf {\bibinfo {volume} {127}},\ \bibinfo {pages} {277201} (\bibinfo {year} {2021})}\BibitemShut {NoStop}%
\bibitem [{\citenamefont {Liu}\ \emph {et~al.}(2021)\citenamefont {Liu}, \citenamefont {Zhao}, \citenamefont {Huang}, \citenamefont {Wu}, \citenamefont {Sheng}, \citenamefont {Xiao},\ and\ \citenamefont {Yang}}]{PhysRevLett.127.277202}%
  \BibitemOpen
  \bibfield  {author} {\bibinfo {author} {\bibfnamefont {H.}~\bibnamefont {Liu}}, \bibinfo {author} {\bibfnamefont {J.}~\bibnamefont {Zhao}}, \bibinfo {author} {\bibfnamefont {Y.-X.}\ \bibnamefont {Huang}}, \bibinfo {author} {\bibfnamefont {W.}~\bibnamefont {Wu}}, \bibinfo {author} {\bibfnamefont {X.-L.}\ \bibnamefont {Sheng}}, \bibinfo {author} {\bibfnamefont {C.}~\bibnamefont {Xiao}},\ and\ \bibinfo {author} {\bibfnamefont {S.~A.}\ \bibnamefont {Yang}},\ }\href@noop {} {\bibfield  {journal} {\bibinfo  {journal} {Phys. Rev. Lett.}\ }\textbf {\bibinfo {volume} {127}},\ \bibinfo {pages} {277202} (\bibinfo {year} {2021})}\BibitemShut {NoStop}%
\bibitem [{\citenamefont {Kirikoshi}\ and\ \citenamefont {Hayami}(2023)}]{PhysRevB.107.155109}%
  \BibitemOpen
  \bibfield  {author} {\bibinfo {author} {\bibfnamefont {A.}~\bibnamefont {Kirikoshi}}\ and\ \bibinfo {author} {\bibfnamefont {S.}~\bibnamefont {Hayami}},\ }\href@noop {} {\bibfield  {journal} {\bibinfo  {journal} {Phys. Rev. B}\ }\textbf {\bibinfo {volume} {107}},\ \bibinfo {pages} {155109} (\bibinfo {year} {2023})}\BibitemShut {NoStop}%
\bibitem [{\citenamefont {Ota}\ \emph {et~al.}(2024)\citenamefont {Ota}, \citenamefont {Shimozawa}, \citenamefont {Muroya}, \citenamefont {Miyamoto}, \citenamefont {Hosoi}, \citenamefont {Nakamura}, \citenamefont {Homma}, \citenamefont {Honda}, \citenamefont {Aoki},\ and\ \citenamefont {Izawa}}]{ota2022zerofield}%
  \BibitemOpen
  \bibfield  {author} {\bibinfo {author} {\bibfnamefont {K.}~\bibnamefont {Ota}}, \bibinfo {author} {\bibfnamefont {M.}~\bibnamefont {Shimozawa}}, \bibinfo {author} {\bibfnamefont {T.}~\bibnamefont {Muroya}}, \bibinfo {author} {\bibfnamefont {T.}~\bibnamefont {Miyamoto}}, \bibinfo {author} {\bibfnamefont {S.}~\bibnamefont {Hosoi}}, \bibinfo {author} {\bibfnamefont {A.}~\bibnamefont {Nakamura}}, \bibinfo {author} {\bibfnamefont {Y.}~\bibnamefont {Homma}}, \bibinfo {author} {\bibfnamefont {F.}~\bibnamefont {Honda}}, \bibinfo {author} {\bibfnamefont {D.}~\bibnamefont {Aoki}},\ and\ \bibinfo {author} {\bibfnamefont {K.}~\bibnamefont {Izawa}},\ }\href@noop {} {\bibfield  {journal} {\bibinfo  {journal} {arXiv:2205.05555}\ } (\bibinfo {year} {2024})}\BibitemShut {NoStop}%
\bibitem [{\citenamefont {Zhao}\ \emph {et~al.}(2020)\citenamefont {Zhao}, \citenamefont {Deng}, \citenamefont {Chen}, \citenamefont {Ross}, \citenamefont {Pet{\v r}{\'\i}{\v c}ek}, \citenamefont {G{\"u}nther}, \citenamefont {Russina}, \citenamefont {Hutanu},\ and\ \citenamefont {Gegenwart}}]{science.aaw1666}%
  \BibitemOpen
  \bibfield  {author} {\bibinfo {author} {\bibfnamefont {K.}~\bibnamefont {Zhao}}, \bibinfo {author} {\bibfnamefont {H.}~\bibnamefont {Deng}}, \bibinfo {author} {\bibfnamefont {H.}~\bibnamefont {Chen}}, \bibinfo {author} {\bibfnamefont {K.~A.}\ \bibnamefont {Ross}}, \bibinfo {author} {\bibfnamefont {V.}~\bibnamefont {Pet{\v r}{\'\i}{\v c}ek}}, \bibinfo {author} {\bibfnamefont {G.}~\bibnamefont {G{\"u}nther}}, \bibinfo {author} {\bibfnamefont {M.}~\bibnamefont {Russina}}, \bibinfo {author} {\bibfnamefont {V.}~\bibnamefont {Hutanu}},\ and\ \bibinfo {author} {\bibfnamefont {P.}~\bibnamefont {Gegenwart}},\ }\href@noop {} {\bibfield  {journal} {\bibinfo  {journal} {Science}\ }\textbf {\bibinfo {volume} {367}},\ \bibinfo {pages} {1218} (\bibinfo {year} {2020})}\BibitemShut {NoStop}%
\bibitem [{\citenamefont {Yatsushiro}\ \emph {et~al.}(2021)\citenamefont {Yatsushiro}, \citenamefont {Kusunose},\ and\ \citenamefont {Hayami}}]{PhysRevB.104.054412}%
  \BibitemOpen
  \bibfield  {author} {\bibinfo {author} {\bibfnamefont {M.}~\bibnamefont {Yatsushiro}}, \bibinfo {author} {\bibfnamefont {H.}~\bibnamefont {Kusunose}},\ and\ \bibinfo {author} {\bibfnamefont {S.}~\bibnamefont {Hayami}},\ }\href@noop {} {\bibfield  {journal} {\bibinfo  {journal} {Phys. Rev. B}\ }\textbf {\bibinfo {volume} {104}},\ \bibinfo {pages} {054412} (\bibinfo {year} {2021})}\BibitemShut {NoStop}%
\bibitem [{\citenamefont {Hayami}\ and\ \citenamefont {Kusunose}(2024)}]{JPSJ.93.072001}%
  \BibitemOpen
  \bibfield  {author} {\bibinfo {author} {\bibfnamefont {S.}~\bibnamefont {Hayami}}\ and\ \bibinfo {author} {\bibfnamefont {H.}~\bibnamefont {Kusunose}},\ }\href@noop {} {\bibfield  {journal} {\bibinfo  {journal} {J. Phys. Soc. Jpn.}\ }\textbf {\bibinfo {volume} {93}},\ \bibinfo {pages} {072001} (\bibinfo {year} {2024})}\BibitemShut {NoStop}%
\bibitem [{\citenamefont {Kusunose}\ \emph {et~al.}(2023)\citenamefont {Kusunose}, \citenamefont {Oiwa},\ and\ \citenamefont {Hayami}}]{PhysRevB.107.195118}%
  \BibitemOpen
  \bibfield  {author} {\bibinfo {author} {\bibfnamefont {H.}~\bibnamefont {Kusunose}}, \bibinfo {author} {\bibfnamefont {R.}~\bibnamefont {Oiwa}},\ and\ \bibinfo {author} {\bibfnamefont {S.}~\bibnamefont {Hayami}},\ }\href@noop {} {\bibfield  {journal} {\bibinfo  {journal} {Phys. Rev. B}\ }\textbf {\bibinfo {volume} {107}},\ \bibinfo {pages} {195118} (\bibinfo {year} {2023})}\BibitemShut {NoStop}%
\bibitem [{\citenamefont {Yanagi}\ \emph {et~al.}(2023)\citenamefont {Yanagi}, \citenamefont {Kusunose}, \citenamefont {Nomoto}, \citenamefont {Arita},\ and\ \citenamefont {Suzuki}}]{PhysRevB.107.014407}%
  \BibitemOpen
  \bibfield  {author} {\bibinfo {author} {\bibfnamefont {Y.}~\bibnamefont {Yanagi}}, \bibinfo {author} {\bibfnamefont {H.}~\bibnamefont {Kusunose}}, \bibinfo {author} {\bibfnamefont {T.}~\bibnamefont {Nomoto}}, \bibinfo {author} {\bibfnamefont {R.}~\bibnamefont {Arita}},\ and\ \bibinfo {author} {\bibfnamefont {M.-T.}\ \bibnamefont {Suzuki}},\ }\href@noop {} {\bibfield  {journal} {\bibinfo  {journal} {Phys. Rev. B}\ }\textbf {\bibinfo {volume} {107}},\ \bibinfo {pages} {014407} (\bibinfo {year} {2023})}\BibitemShut {NoStop}%
\bibitem [{Sup()}]{Supplemental_Material}%
  \BibitemOpen
  \href@noop {} {}\bibinfo {note} {See Supplemental Material, which includes Refs.~\cite{PhysRevB.107.195118,PhysRevB.104.054412,PhysRevB.107.014407,science.aaw1666}, for the symmetry-adapted multipole basis set, the local model Hamiltonian, the distortion parameter dependence of the sublattices, and the filling and molecular field dependence of the ME tensor.}\BibitemShut {Stop}%
\bibitem [{miy()}]{miyamoto}%
  \BibitemOpen
  \href@noop {} {}\bibinfo {note} {T. Miyamoto, S. Hosoi, M. Yatsushiro, K. Ota, K. Izawa, Y. Onuki, D. Aoki, S. Hayami, and M. Shimozawa (unpublished)}\BibitemShut {NoStop}%
\bibitem [{\citenamefont {Oiwa}\ and\ \citenamefont {Kusunose}(2022)}]{JPSJ.91.014701}%
  \BibitemOpen
  \bibfield  {author} {\bibinfo {author} {\bibfnamefont {R.}~\bibnamefont {Oiwa}}\ and\ \bibinfo {author} {\bibfnamefont {H.}~\bibnamefont {Kusunose}},\ }\href@noop {} {\bibfield  {journal} {\bibinfo  {journal} {J. Phys. Soc. Jpn.}\ }\textbf {\bibinfo {volume} {91}},\ \bibinfo {pages} {014701} (\bibinfo {year} {2022})}\BibitemShut {NoStop}%
\bibitem [{\citenamefont {Hayami}\ \emph {et~al.}(2018)\citenamefont {Hayami}, \citenamefont {Yatsushiro}, \citenamefont {Yanagi},\ and\ \citenamefont {Kusunose}}]{PhysRevB.98.165110}%
  \BibitemOpen
  \bibfield  {author} {\bibinfo {author} {\bibfnamefont {S.}~\bibnamefont {Hayami}}, \bibinfo {author} {\bibfnamefont {M.}~\bibnamefont {Yatsushiro}}, \bibinfo {author} {\bibfnamefont {Y.}~\bibnamefont {Yanagi}},\ and\ \bibinfo {author} {\bibfnamefont {H.}~\bibnamefont {Kusunose}},\ }\href@noop {} {\bibfield  {journal} {\bibinfo  {journal} {Phys. Rev. B}\ }\textbf {\bibinfo {volume} {98}},\ \bibinfo {pages} {165110} (\bibinfo {year} {2018})}\BibitemShut {NoStop}%
\bibitem [{\citenamefont {Watanabe}\ and\ \citenamefont {Yanase}(2018)}]{PhysRevB.98.245129}%
  \BibitemOpen
  \bibfield  {author} {\bibinfo {author} {\bibfnamefont {H.}~\bibnamefont {Watanabe}}\ and\ \bibinfo {author} {\bibfnamefont {Y.}~\bibnamefont {Yanase}},\ }\href@noop {} {\bibfield  {journal} {\bibinfo  {journal} {Phys. Rev. B}\ }\textbf {\bibinfo {volume} {98}},\ \bibinfo {pages} {245129} (\bibinfo {year} {2018})}\BibitemShut {NoStop}%
\end{thebibliography}%

\newpage

\begin{widetext}
\begin{center}
  \large\bf
  Supplementary Materials
\end{center}

\section*{S1. Symmetry-adapted multipole basis set for nine-sublattice system}

Here, we show how to construct the symmetry-adapted multipole bases (SAMBs) for the magnetic structure of the nine-sublattice system.
To this end, first, we construct the charge density wave state to describe the charge configuration of a $\sqrt{3}\times\sqrt{3}$ unit cell.
The SAMBs for the charge distribution in the three-sublattice system are given by
\begin{equation}
  \ket{Q_{01}}=\frac{1}{\sqrt{3}}(1,1,1),
  \ket{Q_{02}}=\frac{1}{\sqrt{6}}(-1,2,-1),
  \ket{Q_{03}}=\frac{1}{\sqrt{2}}(-1,0,1),
  \label{eq:charge_unit_cell}
\end{equation}
where the basis is given by $(q_{\mathrm{A}},q_{\mathrm{B}},q_{\mathrm{C}})$ with $q_{i}$ (the charge at site $i$).
All SAMBs in Eq.~\eqref{eq:charge_unit_cell} are orthonormalized satisfying $\braket{Q_{i}|Q_{i}}=\delta_{ij}$.
The schematic picture is shown in Fig.~\ref{fig:charge_distribution}(a).
Using these bases, we introduce the following charge configurations for a $\sqrt{3}\times\sqrt{3}$ unit cell:
\begin{subequations}
  \label{eq:charge_density_wave}
  \begin{equation}
    \mathbb{Q}_{0}^{(0)}=\ket{Q_{01}},
  \end{equation}
  \begin{equation}
    \mathbb{Q}_{0}^{(1)}
    =\sqrt{2}\left[\sqrt{\frac{1}{3}}\ket{Q_{01}}\sin{\tilde{q}_{-}}
    +\sqrt{\frac{2}{3}}\left(\frac{1}{2}\ket{Q_{02}}+\frac{\sqrt{3}}{2}\ket{Q_{03}}\right)\sin{\tilde{q}_{+}}\right],
  \end{equation}
  \begin{equation}
    \mathbb{Q}_{0}^{(2)}
    =-\sqrt{2}\left[\sqrt{\frac{1}{3}}\ket{Q_{01}}\cos{\tilde{q}_{-}}
    +\sqrt{\frac{2}{3}}\left(\frac{1}{2}\ket{Q_{02}}+\frac{\sqrt{3}}{2}\ket{Q_{03}}\right)\cos{\tilde{q}_{+}}\right],
  \end{equation}
\end{subequations}
where $\tilde{q}_{\pm}=\bm{q}\cdot\bm{r} \pm \pi/6$ for $\bm{q}=(1/3,1/3,0)$.
The corresponding charge configurations are shown in Fig.~\ref{fig:charge_distribution}(b).
These charge density wave states belong to the totally symmetric represenation $\mathrm{A}_{1}^{\prime}$ under $D_{\mathrm{3h}}$.
$\mathbb{Q}_{0}^{(0)}$ is the uniform charge distribution, whereas $\mathbb{Q}_{0}^{(1)}$ and $\mathbb{Q}_{0}^{(2)}$ represent the other independent charge density waves with $\bm{q}=(1/3,1/3,0)$.
\begin{figure}[htbp]
  \centering
  \includegraphics[width=\linewidth]{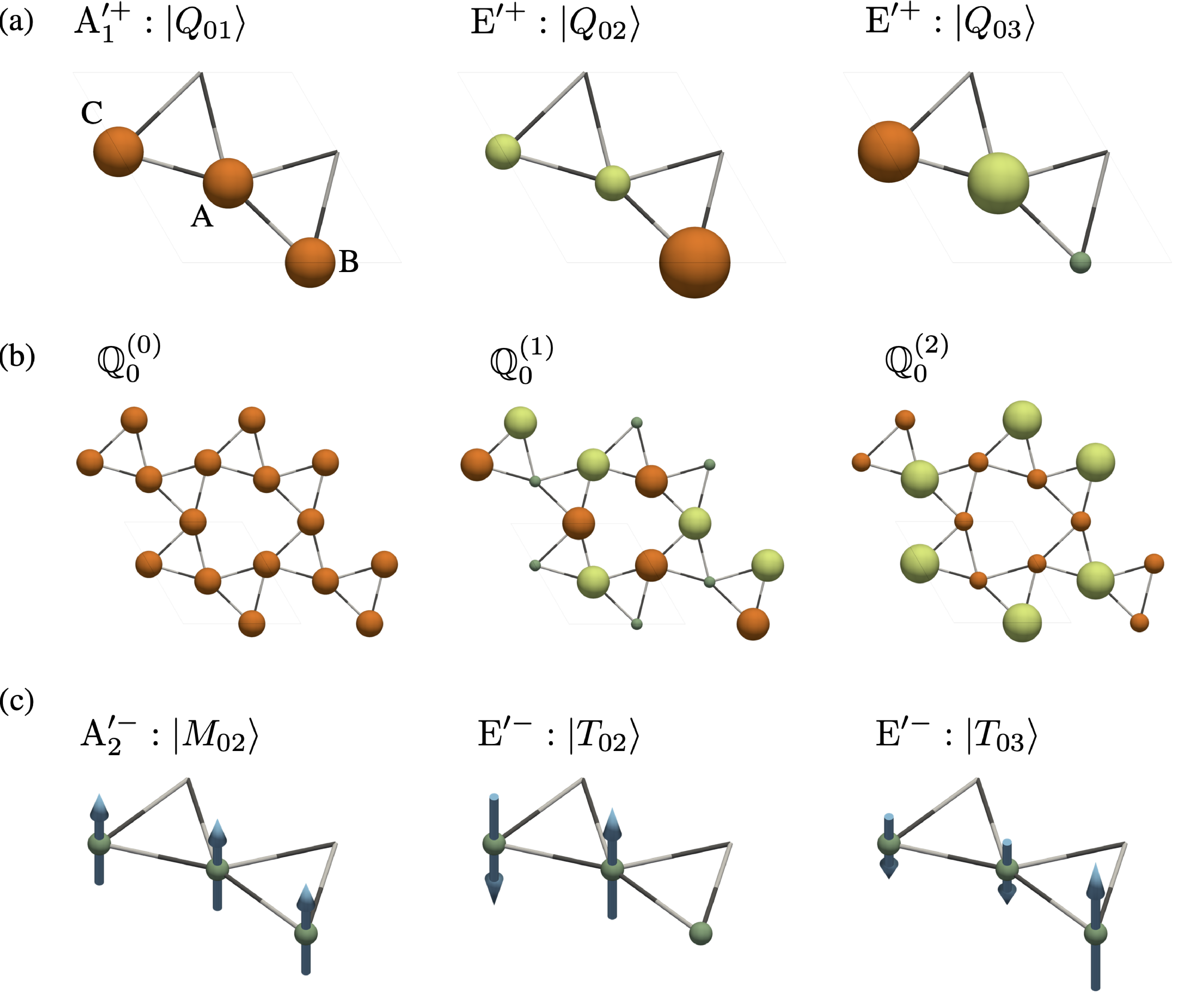}
  \caption{(a)The charge distributions $(q_{\mathrm{A}},q_{\mathrm{B}},q_{\mathrm{C}})$ for the three-sublattice unit cell.
  $\mathrm{A}_{1}^{\prime}$ and $\mathrm{E}^{\prime}$ are the irreducible representations under $D_{\mathrm{3h}}$, where the superscript ``$+$'' represents even parity for the time-reversal operation.
  (b)The charge density wave states with $\bm{q}=(1/3,1/3,0)$ given in Eq.~\eqref{eq:charge_density_wave}.
  (c)SAMBs and their irreducible representation for the three-sublattice out-of-plane magnetic strcture.}
  \label{fig:charge_distribution}
\end{figure}

Next, we construct the SAMBs for the three-sublattice magnetic structures.
The generation of the SAMBs was performed using the open-source software MultiPie~\cite{PhysRevB.107.195118}, and we obtained six bases for the in-plane magnetic structures $\{\ket{M_{01}},\ket{T_{01}},\ket{M_{03}},\ket{M_{04}},\ket{M_{05}},\ket{M_{06}}\}$ in Fig.~2 in the main text and three bases for the out-of-plane magnetic structures $\{\ket{M_{03}},\ket{T_{02}},\ket{T_{03}}\}$ in Fig.~\ref{fig:charge_distribution}(c).

The SAMB for the magentic structure in the $\sqrt{3}\times\sqrt{3}$ unit cell can be constructed by the direct product of Eq.~\eqref{eq:charge_density_wave} and the magnetic structure shown in Fig.~2 in the main text and Fig.~\ref{fig:charge_distribution}(c).
In particular, the spin density wave states of the in-plane magnetic structure are given as follows:
\begin{subequations}
  \label{eq:spin_density_wave}
  \begin{equation}
    \mathbb{M}_{0}^{(1)}=
    \sqrt{\frac{2}{3}}(\ket{M_{01}}\sin{\tilde{q}_{-}}+\ket{M_{01}^{\prime}}\sin{\tilde{q}_{+}}),
  \end{equation}
  \begin{equation}
    \mathbb{M}_{0}^{(2)}=
    -\sqrt{\frac{2}{3}}(\ket{M_{01}}\cos{\tilde{q}_{-}}+\ket{M_{01}^{\prime}}\cos{\tilde{q}_{+}})
  \end{equation}
  with $\ket{M_{01}^{\prime}}=\frac{1}{2}(\ket{M_{04}}+\ket{M_{06}})-\frac{\sqrt{3}}{2}(\ket{M_{03}}+\ket{M_{05}})$,
  \begin{equation}
    \mathbb{T}_{1}^{(1)}=
    \sqrt{\frac{2}{3}}(\ket{T_{01}}\sin{\tilde{q}_{-}}+\ket{T_{01}^{\prime}}\sin{\tilde{q}_{+}}),
  \end{equation}
  \begin{equation}
    \mathbb{T}_{1}^{(2)}=
    -\sqrt{\frac{2}{3}}(\ket{T_{01}}\cos{\tilde{q}_{-}}+\ket{T_{01}^{\prime}}\cos{\tilde{q}_{+}})
  \end{equation}
  with $\ket{T_{01}^{\prime}}=\frac{1}{2}(\ket{M_{03}}-\ket{M_{05}})+\frac{\sqrt{3}}{2}(\ket{M_{04}}-\ket{M_{06}})$,
  \begin{equation}
    \mathbb{M}_{1,0}^{(1)}=
    \sqrt{\frac{2}{3}}(\ket{M_{03}}\sin{\tilde{q}_{-}}+\ket{M_{03}^{\prime}}\sin{\tilde{q}_{+}}),
  \end{equation}
  \begin{equation}
    \mathbb{M}_{1,0}^{(2)}=
    -\sqrt{\frac{2}{3}}(\ket{M_{03}}\cos{\tilde{q}_{-}}+\ket{M_{03}^{\prime}}\cos{\tilde{q}_{+}})
  \end{equation}
  with $\ket{M_{03}^{\prime}}=\frac{1}{2}(\ket{T_{01}}-\ket{M_{05}})-\frac{\sqrt{3}}{2}(\ket{M_{01}}-\ket{M_{06}})$,
  \begin{equation}
    \mathbb{M}_{1,1}^{(1)}=
    \sqrt{\frac{2}{3}}(\ket{M_{04}}\sin{\tilde{q}_{-}}+\ket{M_{04}^{\prime}}\sin{\tilde{q}_{+}}),
  \end{equation}
  \begin{equation}
    \mathbb{M}_{1,1}^{(2)}=
    -\sqrt{\frac{2}{3}}(\ket{M_{04}}\cos{\tilde{q}_{-}}+\ket{M_{04}^{\prime}}\cos{\tilde{q}_{+}})
  \end{equation}
  with $\ket{M_{04}^{\prime}}=\frac{1}{2}(\ket{M_{01}}+\ket{M_{06}})+\frac{\sqrt{3}}{2}(\ket{T_{01}}+\ket{M_{05}})$,
  \begin{equation}
    \mathbb{M}_{2,0}^{(1)}=
    \sqrt{\frac{2}{3}}(\ket{M_{05}}\sin{\tilde{q}_{-}}-\ket{M_{05}^{\prime}}\sin{\tilde{q}_{+}}),
  \end{equation}
  \begin{equation}
    \mathbb{M}_{2,0}^{(2)}=
    -\sqrt{\frac{2}{3}}(\ket{M_{05}}\cos{\tilde{q}_{-}}-\ket{M_{05}^{\prime}}\cos{\tilde{q}_{+}})
  \end{equation}
  with $\ket{M_{05}^{\prime}}=\frac{1}{2}(\ket{T_{01}}+\ket{M_{03}})+\frac{\sqrt{3}}{2}(\ket{M_{01}}-\ket{M_{04}})$,
  \begin{equation}
    \mathbb{M}_{2,1}^{(1)}=
    \sqrt{\frac{2}{3}}(\ket{M_{06}}\sin{\tilde{q}_{-}}+\ket{M_{06}^{\prime}}\sin{\tilde{q}_{+}}),
  \end{equation}
  \begin{equation}
    \mathbb{M}_{2,1}^{(2)}=
    -\sqrt{\frac{2}{3}}(\ket{M_{06}}\cos{\tilde{q}_{-}}+\ket{M_{06}^{\prime}}\cos{\tilde{q}_{+}})
  \end{equation}
  with $\ket{M_{06}^{\prime}}=\frac{1}{2}(\ket{M_{01}}+\ket{M_{04}})-\frac{\sqrt{3}}{2}(\ket{T_{01}}-\ket{M_{03}})$.
\end{subequations}
\begin{figure}[htbp]
  \centering
  \includegraphics[width=\linewidth]{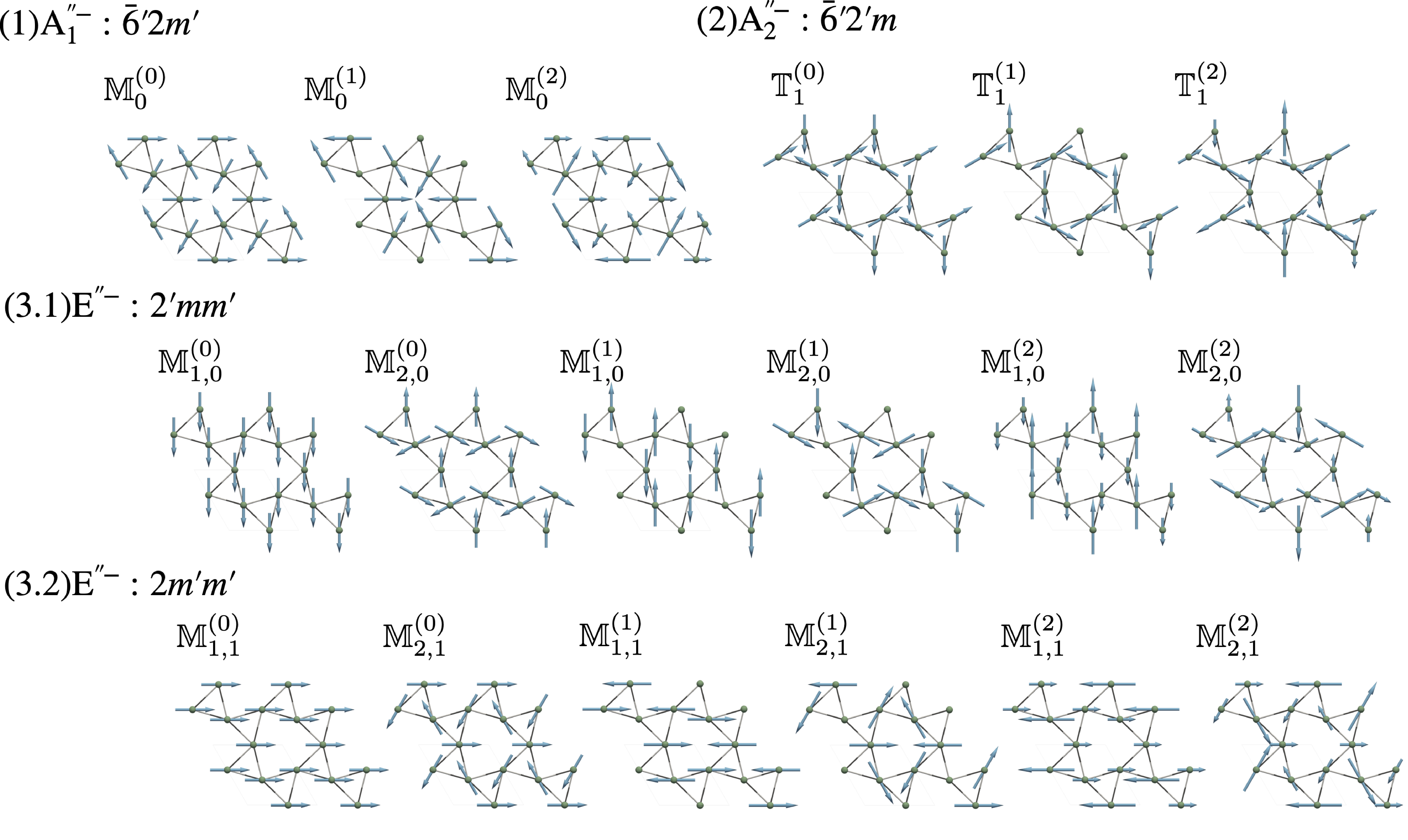}
  \caption{\label{fig:SAMB_9sub_mag}The symmetry-adapted bases for the $\sqrt{3}\times\sqrt{3}$ unit cell (nine-sublattice) with the corresponding irreducible representation under $D_{\mathrm{3h}}$.
  We also give the magnetic point group~\cite{PhysRevB.104.054412} under the magnetic ordering.
  The superscripts in multipoles denote the uniform state as $(0)$, whereas the spin-density wave state given in Eq.~\eqref{eq:spin_density_wave} as $(1),(2)$.
  $(\mathbb{M}_{i,0}^{(j)},\mathbb{M}_{i,1}^{(j)})$ is the basis set of the two-dimensional irreducible representation $\mathrm{E}^{\prime\prime}$.
  }
\end{figure}
Figure~\ref{fig:SAMB_9sub_mag} shows the SAMB set for describing any in-plane magnetic structure for the $\sqrt{3}\times\sqrt{3}$ unit cell, each of which is composed of the magnetic cluster multipoles given in Eq.~\eqref{eq:spin_density_wave} plus a three-sublattice uniform state, where we relabelled the uniform state as $\{\ket{M_{01}},\ket{T_{01}},\ket{M_{03}},\ket{M_{04}},\ket{M_{05}},\ket{M_{06}}\}\to\{\mathbb{M}_{0}^{(0)},\mathbb{T}_{1}^{(0)},\mathbb{M}_{1,0}^{(0)},\mathbb{M}_{1,1}^{(0)},\mathbb{M}_{2,0}^{(0)},\mathbb{M}_{2,1}^{(0)}\}$.
All the SAMBs $X,Y=M$ or $T$ are orthonormalized such as $\mathrm{Tr}[XY]=\delta_{XY}$.
We note that the single-$q$ magnetic ordering can be classified in terms of the $\bm{k}$-group at the magnetic ordering vector~\cite{PhysRevB.107.014407}.
Since the $\bm{k}$-point group at $\bm{q}=(1/3,1/3,0)$, the K point in the Brillouin zone, coincides with the one at $\bm{q}=\bm{0}$, the $\Gamma$ point, we can classify the SAMBs in Fig.~\ref{fig:SAMB_9sub_mag} under $D_{\mathrm{3h}}$.

In the main text, the partially ordered state corresponds to $\mathbb{T}_{1}^{(1)}$.
Meanwhile, the kagome spin ice state can be expressed by the superposition of $\mathbb{T}_{1}^{(0)}$ and $\mathbb{T}_{1}^{(2)}$.
Thus, the partially ordered state is characterized by the density wave of the MTD, i.e., the antiferrotoroidal ordering, whereas the kagome spin ice state is characterized by the ferritoroidal ordering.

By performing a similar procedure, we can construct the SAMB for the perfect kagome structure with $\delta=0$ under $D_{\mathrm{6h}}$ symmetry, where the same SAMBs are constructed.
Meanwhile, some SAMBs belong to a different irreducible representation;
for example, $(\mathbb{T}_{1}^{(0)}$ and $\mathbb{T}_{1}^{(2)})$ belong to the $\mathrm{B}_{2g}^{-}$ representation, and $\mathbb{T}_{1}^{(1)}$ belongs to the $\mathrm{A}_{1u}^{-}$ representation in the perfect kagome structure.
Since the MTD $T_{z}$ is activated under $\mathrm{A}_{1u}^{-}$ representation, the partially ordered state with $\mathbb{T}_{1}^{(1)}$ induces the nonzero MTD moment, while the kagome spin ice state with $\mathbb{T}_{1}^{(0)}$ and $\mathbb{T}_{1}^{(2)}$ does not.
Such a symmetry analysis is consistent with the result from the model analysis at $\delta=0$ in the main text.

\section*{S2. Model Hamiltonian}

\subsection{Distortion parameter}

We introduced the distortion parameter $\delta$ in the main text.
This parameter is included in the Hamiltonian through the coordinates of each site, which are given as follows:
\begin{equation}
  \begin{aligned}
    &\, \bm{r}_{\mathrm{A}_{1}}=\frac{1-\delta}{2}\left(\frac{1}{2},\frac{\sqrt{3}}{2}\right),
    \bm{r}_{\mathrm{B}_{1}}=\frac{1-\delta}{2}\left(-1,0\right),
    \bm{r}_{\mathrm{C}_{1}}=\frac{1-\delta}{2}\left(\frac{1}{2},-\frac{\sqrt{3}}{2}\right),
    \\
    &\, \bm{r}_{\mathrm{A}_{2}}=\frac{1+\delta}{2}\left(-\frac{1}{2},-\frac{\sqrt{3}}{2}\right),
    \bm{r}_{\mathrm{B}_{2}}=\frac{1+\delta}{2}\left(1,0\right),
    \bm{r}_{\mathrm{C}_{2}}=\frac{1+\delta}{2}\left(-\frac{1}{2},\frac{\sqrt{3}}{2}\right),
    \\
    &\, \bm{r}_{\mathrm{A}_{3}}=(-1,0)+\bm{r}_{\mathrm{A}_{1}},
    \bm{r}_{\mathrm{B}_{3}}=\left(-\frac{1}{2},\frac{\sqrt{3}}{2}\right)+\bm{r}_{\mathrm{B}_{1}},
    \bm{r}_{\mathrm{C}_{3}}=\left(\frac{1}{2},\frac{\sqrt{3}}{2}\right)+\bm{r}_{\mathrm{C}_{1}}
    \\
  \end{aligned}
\end{equation}
with a lattice constant $a=1$.
As shown in Fig.~\ref{fig:distortion}, as $\delta$ is varied from zero, the lattice continuously changes from a perfect kagome structure to a distorted kagome one.
\begin{figure}[htbp]
  \centering
  \includegraphics[width=\linewidth]{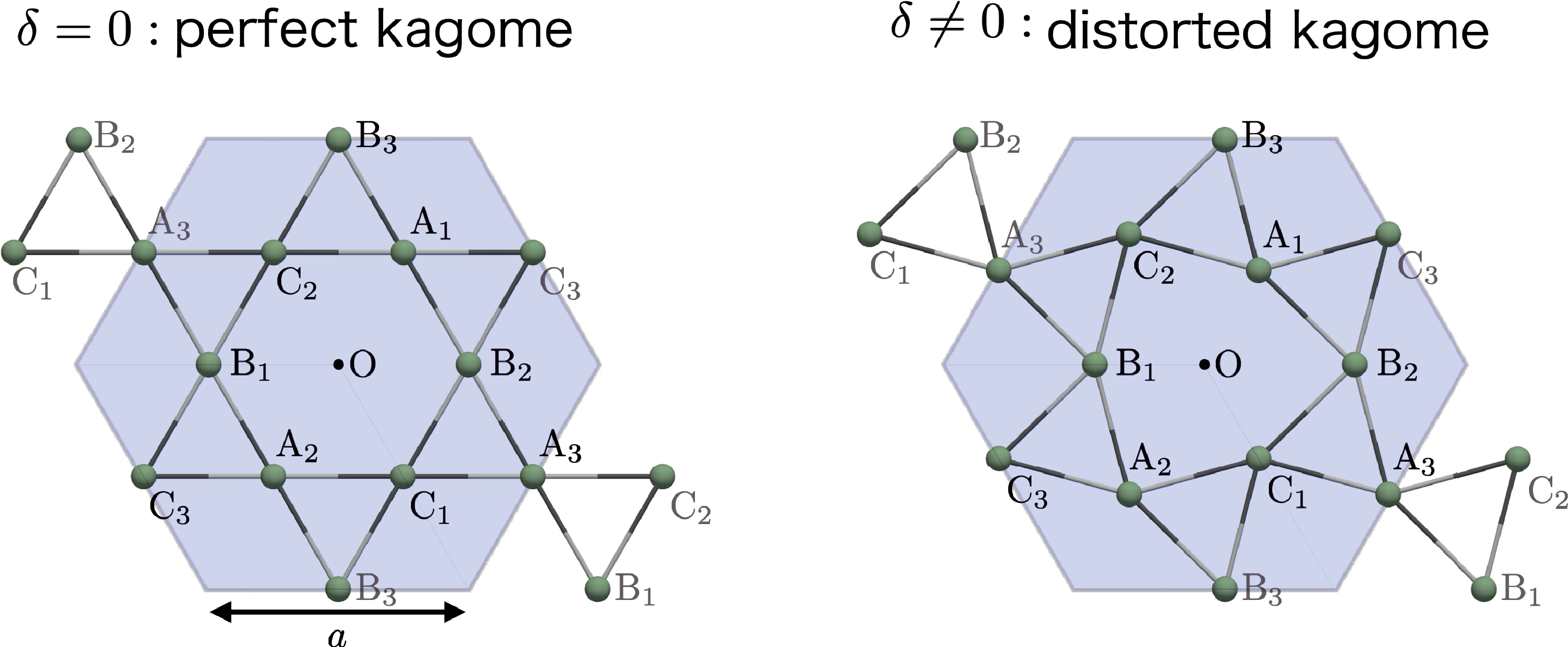}
  \caption{\label{fig:distortion}The relation between the perfect and distorted kagome structures with and without the distortion parameter $\delta$.}  
\end{figure}

\subsection{Local Hamiltonian}

On the sites of the distorted kagome strcuture, sublattice-dependent polarizations are associated with spatial inversion symmetry breaking, as shown in Fig.~\ref{fig:local_model}(a).
To incorporate the anisotropy of polarization, we construct a local Hamiltonian for the three $p$-orbitals $(p_{x},p_{y},p_{z})$.
Since the site symmetry is $C_{\mathrm{2v}}$, the Hamiltonian at site $\mathrm{B}$ (the principal axis $[100]$) is expressed as follows:
\begin{equation}
  h_{\mathrm{loc}}=B_{2}^{0}Q_{20}+B_{2}^{2}Q_{22}+\lambda\bm{l}\cdot\bm{s}.
\end{equation}
The first two terms are the crystalline electric field (CEF) term, where $Q_{20},Q_{22}$ are the electric quadrupole operators whose matrix elements are 
\begin{equation}
  Q_{20}=\frac{1}{\sqrt{6}}
  \begin{pmatrix}
    -1 & 0 & 0 
    \\
    0 & -1 & 0 
    \\
    0 & 0 & 2 
    \\
  \end{pmatrix},
  Q_{22}=\frac{1}{\sqrt{2}}
  \begin{pmatrix}
    1 & 0 & 0 
    \\
    0 & -1 & 0 
    \\
    0 & 0 & 0 
    \\
  \end{pmatrix},
\end{equation}
with the corresponding parameters $B_{2}^{0},B_{2}^{2}$.
The last term is the atomic spin--orbit coupling (SOC) term, where $\bm{l}$ and $\bm{s}$ are the orbital and spin angular momentum operators, and $\lambda$ is a coupling constant.
The local Hamiltonian at the other sublattices can be constructed by considering the three-fold rotational operation in orbital--spin space.
\begin{figure}[htbp]
  \centering
  \includegraphics[width=\linewidth]{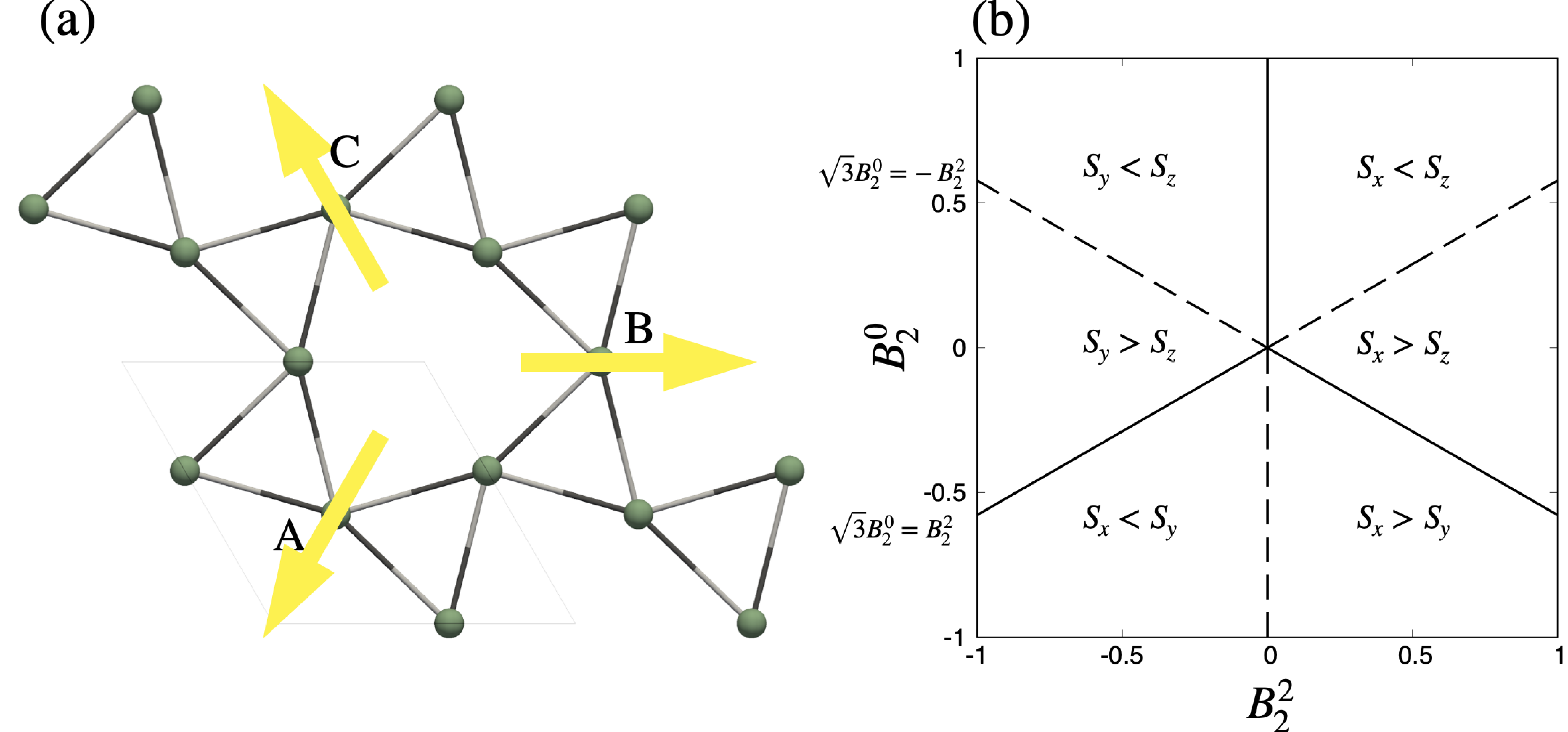}
  \caption{\label{fig:local_model}
  (a)The polarization direction at each site.
  (b)The relation between the CEF parameters and the easy axes of the magnetic moment based on the effective spin Hamiltonian in Eq.~\eqref{eq:effective_spin_Hamiltonian} at site $\mathrm{B}$.
  $S_{i}>S_{j}$ means that the $i$-component of the magnetic moment is the primary easy axis and the $j$-component is the secondary one.
  Crossing the solid(dashed) line represents the change of the easy axis with(without) the change of the ground-state CEF level in the perturbed region.
  }  
\end{figure}

We examine the magnetic anisotropy by performing the perturbation analysis.
Within the second-order perturbation in terms of $\lambda$, an effective spin Hamiltonian is derived as 
\begin{equation}
  h_{\mathrm{eff}}=-\lambda^{2}\sum_{\alpha\beta}\Lambda_{\alpha\beta}\sigma_{\alpha}\sigma_{\beta}
  \label{eq:effective_spin_Hamiltonian}
\end{equation}
with 
\begin{equation*}
  \Lambda_{\alpha\beta}=\sum_{e}\frac{\langle{g|l_{\alpha}|e}\rangle\langle{e|l_{\beta}|g}\rangle}{E_{e}-E_{g}}.
\end{equation*}
Here, $\sigma_{\alpha}(\alpha=x,y,z)$ is the Pauli matrix in spin space.
$E_{g}(E_{e})$ is the ground state (excited state) energy, and $\ket{g}(\ket{e})$ is the corresponding eigenstate.
We note that $\Lambda_{\alpha\beta}=\delta_{\alpha\beta}\Lambda_{\alpha}$ under the present symmetry.
Figure~\ref{fig:local_model}(b) shows the relations between the CEF parameters and the easy axes at site $\mathrm{B}$ based on Eq.~\eqref{eq:effective_spin_Hamiltonian}.
According to Ref.~\cite{science.aaw1666}, the magnetic moment of $\mathrm{Ho}$ is constrained in the $xy$-plane.
Since the easy axis at site $\mathrm{B}$ is the $y$-axis, we choose the CEF parameters in the $S_{x}<S_{y}$ region in Fig.~\ref{fig:local_model}(b).
At $\lambda=0$, the ground-state Kramers pair corresponds to the $p_{z}$-orbital.
The effect of SOC brings hybridization between $p_{x},p_{y}$-orbitals, and $p_{z}$-orbital, which is important for induced the $sp$-orbital hybridization.
In the end, we chose the ground-state Kramers pair as the $p$-orbital bases of the effective model in the main text by setting $B_{2}^{0}=-0.6/\sqrt{3},B_{2}^{2}=-0.5$, and $\lambda=1$.

\subsection{Model parameter dependence of ME tensor}

In this subsection, we show the model parameter dependence of the ME tensor.
Figures~\ref{fig:ME_distribution}(a) and (b) show the antisymmetric component of the ME tensor $\chi_{xy}$ on the distorted kagome structure in the plane of the electron filling $n$ and the molecular field $h$. 
The three left panels in each figure show the data of the site-resolved ME tensor, while the rightmost one shows that of the total ME tensor.
In the partially ordered state, the total ME tensor is suppressed in the low-filling region, which means that the antiferro MT nature is dominant.
On the other hand, in the kagome spin ice state, the ME tensor exhibits the ferri MT nature near $n\simeq 2.5$.
\begin{figure}[htbp]
  \centering
  \includegraphics[width=\linewidth]{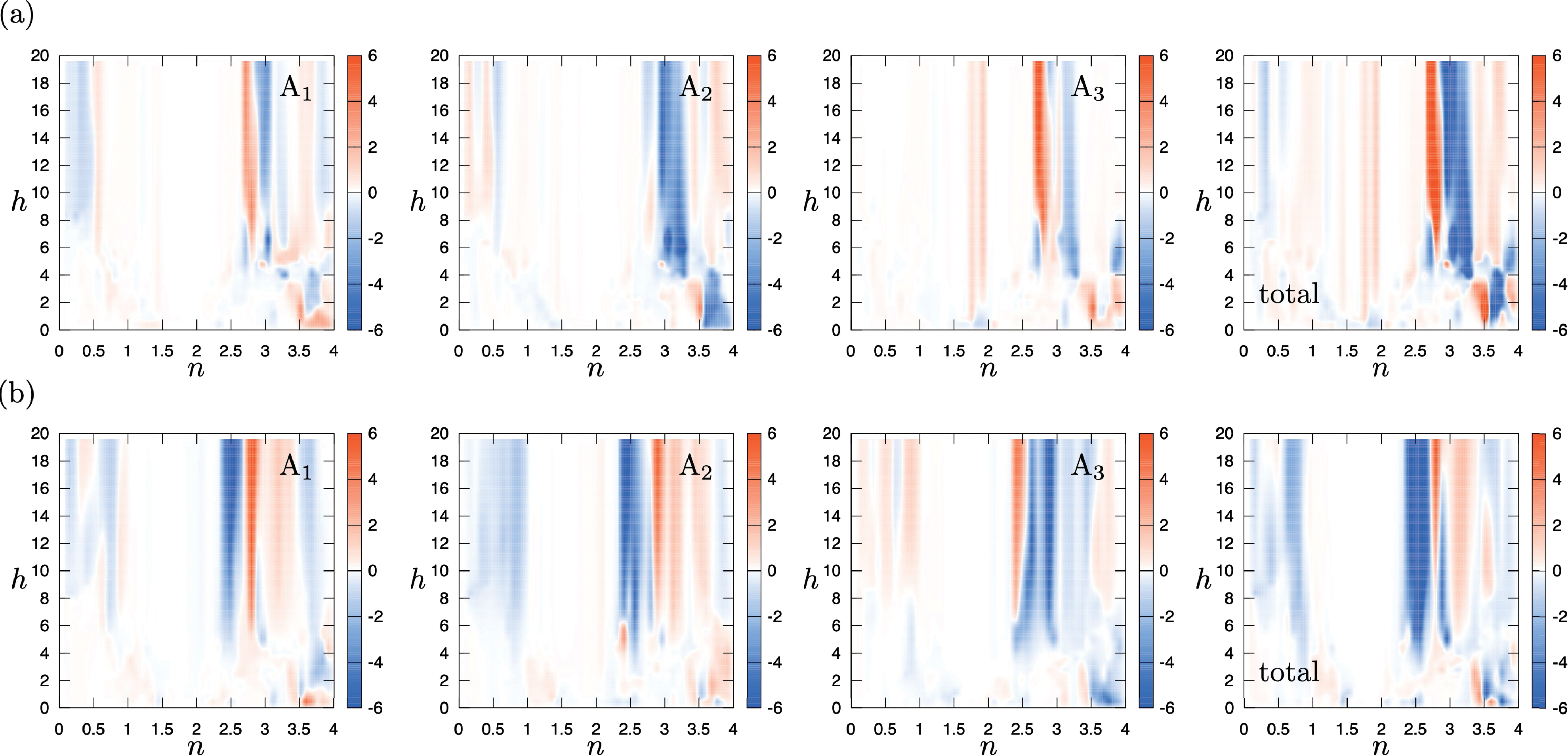}
  \caption{\label{fig:ME_distribution}The contour plot of the antisymmetric component of the ME tensor $\chi_{xy}$ in the filling $n$ and molecular field $h$ plane under (a) the partially ordered state and (b) the kagome spin ice state with $\delta=0.16$.}
\end{figure}
We note, however, that the ME tensor shows a deviation from the above correspondence between the partially ordered state and the antiferro MT  nature (the kagome spin ice state and the ferri MT nature) in some model-parameter regions, which might be attributed to complicated multiband contributions in the ME tensor.

Figure~\ref{fig:ME_distortion_2} shows the distortion parameter dependence for the different $(n,h)$ parameters than in the main text.
In the partially ordered state, $\chi_{xy}^{(\mathrm{A})}$ exhibits the ferri MT-like behavior even in the distorted kagome structure as in the perfect kagome structure $\delta=0$, as shown in Fig.~\ref{fig:ME_distortion_2}(a).
However, we note that, even in this parameter, the even-order $O(\delta^{2n})$ contributes as $\chi_{xy}^{(\mathrm{A})\mathrm{A}_{1}}=\chi_{xy}^{(\mathrm{A})\mathrm{A}_{2}}\neq 0,\chi_{xy}^{(\mathrm{A})\mathrm{A}_{3}}\neq 0$, whereas the odd-order $O(\delta^{2n+1})$ as $\chi_{xy}^{(\mathrm{A})\mathrm{A}_{1}}=-\chi_{xy}^{(\mathrm{A})\mathrm{A}_{2}}\neq 0,\chi_{xy}^{(\mathrm{A})\mathrm{A}_{3}}=0$.
On the other hand, the ME tensor in the spin ice state grows gradually with the distortion, as shown in Fig.~\ref{fig:ME_distortion_2}(b).
In this case, $\chi_{xy}^{(\mathrm{A})\mathrm{A}_{1}}>0, \chi_{xy}^{(\mathrm{A})\mathrm{A}_{2}}<0$ remains even at $\delta=0.16$.
These results suggest that the behavior of the ME tensor component remains that of the perfect kagome structure if the even-order contribution $O(\delta^{2n})$ is dominant.

\begin{figure}[htbp]
  \centering
  \includegraphics[width=\linewidth]{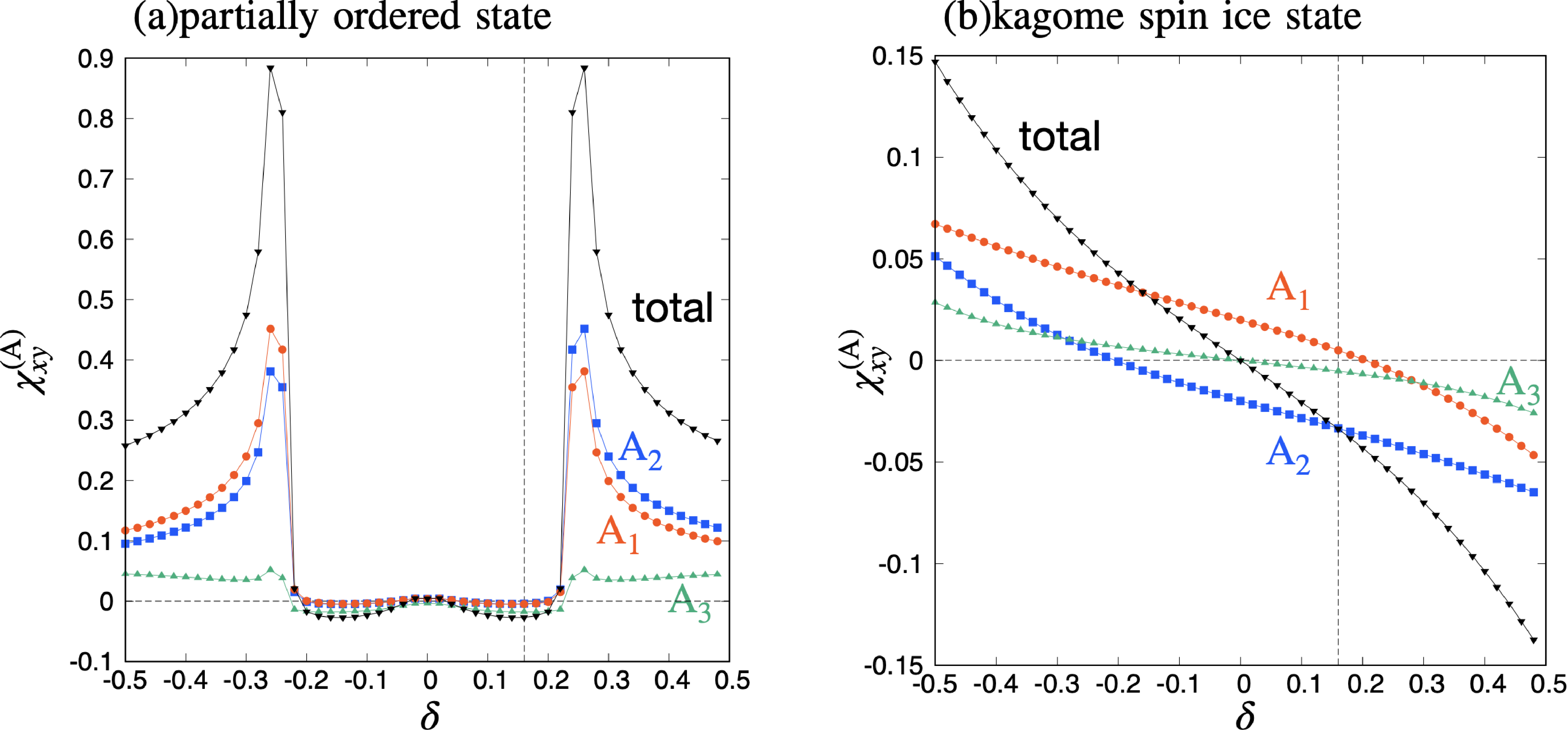}
  \caption{\label{fig:ME_distortion_2}
  Distortion parameter $(\delta)$ dependence of the ME tensor $\chi^{\rm (A)}_{xy}$ per site under (a) the partially ordered state and (b) the kagome spin ice state at $n=1.5$ and $h=15$.
  The vertical dotted line at $\delta=0.16$ indicates the case of HoAgGe.
  The other parameters are the same as Fig.~5 in the main text.}
\end{figure}

\end{widetext}

\end{document}